\documentclass[12pt]{iopart}
\usepackage{iopams} 

\usepackage{eurosym}
\usepackage{eurosym}
\usepackage{amssymb}
\usepackage{graphicx}
\usepackage{hyperref}

\begin{document}

\title{Quantum correlations and coherence between two particles interacting
with a thermal bath}
\author{Marco Nizama$^{1,2}$ and\ Manuel O. C\'{a}ceres$^{1}$}
\address{$^{1}$ Centro At\'{o}mico Bariloche, CNEA, Instituto Balseiro, U. Nac. de Cuyo, and
CONICET, 8400 Bariloche, Argentina.\\
$^{2}$Departamento de F\'{\i}sica, Universidad Nacional del Comahue, 8300,
Neuquen, Argentina} \ead{caceres@cab.cnea.gov.ar}
\date{}

\begin{abstract}
\ \newline
Quantum correlations and coherence generated between two free spinless
particles in the lattice, interacting with a common quantum phonon bath are
studied. The reduced density matrix has been solved in the Markov approach.
We show that the bath induces correlations between the particles. The
coherence induced by the bath is studied calculating: off-diagonal elements
of the density matrix, spatiotemporal dispersion, purity, and quantum mutual
information. We have found a characteristic time-scale pointing out when
this coherence is maximum. In addition a Wigner-like distribution in the
phase-space (lattice) has been introduced as an indirect indicator of the
quantumness of total correlations and coherence induced by the thermal bath.
The negative volume of the Wigner function shows also a behavior which is in
agreement with the time-scale that we have found. A Gaussian distribution
for the profile of particles is not obtained and interference pattern are
observed as the result of bath induced coherence. As the temperature of bath
vanishes the ballistic behavior of the tight-binding model is recovered. The
geometric quantum discord has been calculated to characterize the nature of
the correlations.

\end{abstract}

\maketitle

\section{Introduction}

In quantum systems, damping and fluctuations enter through the coupling with
an external large bath $\mathcal{B}$. The conventional treatment computes
the reduced density matrix of the system, $\mathcal{S}$, by expanding in the
coupling strength to the bath and eliminating these variables. Nevertheless,
some extra approximations must be introduced to arrive to a completely
positive map \cite{Alicki}. This conclusion can be summarized in the
appropriated structure that the Quantum\ Master Equation (QME) must have
after the elimination of bath. In fact, this QME is known to be acceptable
only for few particular cases \cite{Joos,Breuer,vanKampen04,holanda,jpa}.

Here we study a minimal model that indeed leads to a well defined QME
(completely positive infinitesimal semigroup), then we can compute
analytically several quantum measures. In this way we can show that a
thermal bath generates not only dissipation, but indeed induces coherence
and non-classical correlations between particles immersed in it \cite%
{Bath-Induced}. In an analogous way, total correlations generated between a
Spin $\frac{1}{2}$ (the system) and the Magnet apparatus (pointer variable)
coupled to a boson thermal bath show that correlations develop in time,
reach a maximum, then disappear later and later \cite{holanda}. Other
similar works have been proposed to show bath-generated correlations in a
system, induced from a thermal common bath \cite{Bath-Induced}.

In order to study bath-induced correlations, we present exact calculations
on the dynamics of spinless Quantum Walks (QWs) \cite{aharanov,vK95,blumen2}%
. As expected, quantum measures will vanish as time goes on due to the
dissipation. Then, the important point would be to characterize when these
induced correlations and coherences start to decrease. We will present
analytical calculations of these measures and also we compute the
characteristic time-scale when they are maxima before being wiped\ out by
the dissipation. As an indirect measure of the quantum character of the
state of the system, the negative volume of the Wigner function and the
quantum coherence from off-diagonal elements of the density matrix --as a
function of time-- have also been studied.

There are several measures to get information concerning quantum
correlations, in particular, we will focus on the quantum discord \cite%
{Discord}. As we mentioned before, our system presents coherence and
non-classical correlations, in particular, we can indirectly measure the
nonclassical correlation by calculating the geometric quantum discord, which
in fact is an accepted measure despite its criticisms \cite{Modi}. Then we
can use these results to measure the quantum to classical transition, for
example, in qubit systems \cite{nielsen,physA-1,physA-2}.

The simplest implementations that reflect the role of a coherent
superposition can be proposed in the framework of QW experiments, or its
numerical simulations \cite{zahringer,schreiber,broome,schmitz,karski,NMR}.
A Dissipative QW (DQW) has also been defined as a spinless particle moving
in a lattice and interacting with a phonon bath \cite{vK95,kempe,manuel}. In
particular in this work we will implement explicit calculations for a system 
$\mathcal{S}$ constituted by two \textit{distinguishable} particles in a
one-dimensional regular lattice. The present approach can also be extended
to tackle the many-body fermionic particles (or bosonic, see appendix A);
then, pointing out the interplay between particle-particle and bath-particle
interactions.

\section{Dissipative Quantum Walks}

The goal of this section is not intended to derive the completely positive
infinitesimal generator, this section is presented to show explicitly the
contribution responsible of the bath-induced coherence in the system, see
appendix B. Here we introduce shift operators that define the model of two
free distinguishable spinless particles (system $\mathcal{S})$ coupled to a
common phonon bath $\mathcal{B}$. The generalization to bosonic or fermionic
particles can be done in a similar way using Fock's representation, see
appendix A.

The total Hamiltonian for $\mathcal{S}$ coupled to a common bath $\mathcal{B}
$ can be written using the Wannier basis in the following way (for more
details see \cite{vK95,manuel}). Let the total Hamiltonian be 
\begin{equation}
H_{\mathcal{T}}=H_{\mathcal{S}}+H_{\mathcal{B}}+H_{\mathcal{SB}}.  \label{Ht}
\end{equation}%
$H_{\mathcal{S}}$ is the \textit{free} \textit{tight-binding} Hamiltonian
(our system $\mathcal{S}$) 
\begin{equation}
H_{\mathcal{S}}=2E_{0}\mathbf{I}-\frac{\Omega }{2}\left( a_{12}^{\dag
}+a_{12}\right) ,  \label{TH-H}
\end{equation}%
here $\{a_{12}^{\dag },a_{12}\}$ are shift operators for the particles
labeled $1$ and $2$, and $\mathbf{I}$ the identity in the Wannier basis%
\[
\mathbf{I}=\sum_{s,s^{^{\prime }}}\arrowvert s,s^{^{\prime }}\rangle \langle
s,s^{^{\prime }}\arrowvert
\]%
\begin{equation}
a_{12}^{\dag }\arrowvert s_{j},s_{l}\rangle =\arrowvert s_{j}+1,s_{l}\rangle
+\arrowvert s_{j},s_{l}+1\rangle  \label{RR1}
\end{equation}%
\begin{equation}
a_{12}\arrowvert s_{j},s_{l}\rangle =\arrowvert s_{j}-1,s_{l}\rangle +%
\arrowvert s_{j},s_{l}-1\rangle .  \label{RR2}
\end{equation}%
Note that a \textquotedblleft shift operator\textquotedblright\ translates
each particle individually. Here we have used a \textquotedblleft
pair-ordered\textquotedblright\ bra-ket $\arrowvert s_{j},s_{l}\rangle $
representing the particle \textquotedblleft $1$\textquotedblright\ at site $%
s_{j}$ and particle \textquotedblleft $2$\textquotedblright\ at site $s_{l}$%
; note that these operators generate the \textit{free} \textit{tight-binding}
Hamiltonian (\ref{TH-H}), see Appendix A. From Eqs. (\ref{RR1})-(\ref{RR2})
it is simple to see that 
\begin{equation}
\left[ a_{12}^{\dag },a_{12}\right] =0,  \label{RR10}
\end{equation}%
and also that 
\begin{equation}
a_{12}a_{12}^{\dag }\arrowvert s_{j},s_{l}\rangle \!=\!2\arrowvert %
s_{j},s_{l}\rangle +\arrowvert s_{j}-1,s_{l}+1\rangle +\arrowvert %
s_{j}+1,s_{l}-1\rangle .  \label{RR3}
\end{equation}

$H_{\mathcal{B}}$ is the phonon bath $H_{\mathcal{B}}=\sum\limits_{k}\hbar
\omega _{k}\mathcal{B}_{k}^{^{\dag }}\mathcal{B}_{k}$, thus $\{\mathcal{B}%
_{k}^{^{\dag }},\mathcal{B}_{k}\}$ are bosonic operator characterizing the
thermal bath in equilibrium. In eq. (\ref{Ht}) the term $H_{\mathcal{SB}}$
is the interaction Hamiltonian between $\mathcal{S}$ and $\mathcal{B}$,%
\begin{equation}
H_{\mathcal{SB}}=\hbar \Gamma \left( a_{12}\otimes \sum\limits_{k}v_{k}%
\mathcal{B}_{k}+a_{12}^{\dag }\otimes \sum\limits_{k}v_{k}^{\ast }\mathcal{B}%
_{k}^{^{\dag }}\right) ,  \label{Hint}
\end{equation}%
where $v_{k}$ represents the spectral function of the phonon bath, and $%
\Gamma $ is the interaction parameter in the model. This is a minimal
interacting model useful for our purposed study. Getting the QME from this
interaction model produces clearly two separable contributions, this fact
will be studied in detail in the next sections.

To study a non-equilibrium evolution for the system $\mathcal{S}$, we
calculate from (\ref{Ht}) --eliminating the bath variables-- a dissipative
quantum infinitesimal generator (see appendix A in \cite{mm}). Therefore,
tracing out bath variables, and in the Ohmic approximation, we can write the
Markov Quantum Master Equation (QME) \cite{Alicki,holanda,manuel}:

\begin{eqnarray}
\frac{d\rho }{dt}\! &=&\!\frac{-i}{\hbar }\left[ H_{eff},\rho \right] +\frac{%
D}{2}\left( 2a_{12}\rho a_{12}^{\dag }-a_{12}^{\dag }a_{12}\rho -\rho
a_{12}a_{12}^{\dag }\right)  \nonumber \\
&+&\frac{D}{2}\left( 2a_{12}^{\dag }\rho a_{12}-a_{12}a_{12}^{\dag }\rho
-\rho a_{12}^{\dag }a_{12}\right) ,  \label{QME}
\end{eqnarray}%
where $D\equiv \Gamma ^{2}k_{B}T/\hbar $, here $T$ is the temperature of the
bath $\mathcal{B}$. We point out that due to the particular interaction
Hamiltonian model $H_{\mathcal{SB}}$ that we have used, it is possible to
see that the algebra is closed for the operators of $\mathcal{S}$, then we
can prove the QME is a bonafide semigroup \cite{jpa}. Adding $-2E_{0}+\Omega 
$ to $H_{\mathcal{T}}$ the effective Hamiltonian turns to be 
\[
H_{eff}=\Omega \left( \mathbf{I}-\frac{a_{12}^{\dag }+a_{12}}{2}\right)
-\hbar \omega _{c}a_{12}a_{12}^{\dag }, 
\]%
where $\omega _{c}$ is the frequency cut off in the Ohmic approximation.
This Hamiltonian is the natural extension of van Kampen's Hamiltonian for
two spinless particles in the lattice \cite{vK95,mm}.

From the QME (\ref{QME}) we can see that the term%
\begin{equation}
\frac{-D}{2}\left( a_{12}^{\dag }a_{12}\rho +\rho a_{12}a_{12}^{\dag
}+a_{12}a_{12}^{\dag }\rho +\rho a_{12}^{\dag }a_{12}\right) ,
\label{part-bath}
\end{equation}%
is the responsible of generating coherence in the system (see appendix B),
the structure of this operator can be realized from the analysis of
operators like $a_{12}a_{12}^{\dag }$, see (\ref{RR3}). While the
decoherence of the system comes from the term%
\[
\frac{D}{2}\left( 2a_{12}\rho a_{12}^{\dag }+2a_{12}^{\dag }\rho
a_{12}\right) , 
\]%
and its interpretation can be done just in terms of one-step translation of
particles (see appendix A in Fock's representation).

Here we will focus in the highly dissipative regime so we can take $\omega
_{c}=0$ without lost of generality. It can be seen from Eq.(\ref{QME}) that
as $D\rightarrow 0$ the unitary evolution is recovered (the tight-binding
model). The limit $D\rightarrow \infty $ (or $\Omega \rightarrow 0$)
corresponds to the case when the effective Hamiltonian disappears, then we
would expect a \textit{pure random dynamics} corresponding to two Random
Walk (RW). Nevertheless, for the \textit{quantum }two-body problem that we
are working, the classical profile, even when $D/\hbar \Omega \gg 1$, cannot
be reached because coherence have been induced from bath $\mathcal{B}$.

We will solve this QME (\ref{QME}) using a localized Initial Condition (IC)
in the Wannier lattice, i.e.,:%
\begin{equation}
\rho (t=0)=\arrowvert0,0\rangle \langle 0,0\arrowvert\equiv \arrowvert\vec{0}%
\rangle \langle \vec{0}\arrowvert.  \label{IC1}
\end{equation}%
The operational calculus in the QME\ will be done using a two-particles
Wannier vector state to evaluate elements of the density matrix $\rho (t)$.

\section{The Two-Body Solution of QME}

Using (\ref{RR1})-(\ref{RR3}) the QME (\ref{QME}) can be worked out. In
particular to find the analytical solution of $\rho (t)$ is it simpler to do
the calculations in Fourier representation. Then we introduce here the
two-particle Fourier basis (similar calculus where done for the one-particle
problem \cite{manuel,mm,physA-1,physA-2}). The Fourier "\textit{bra-ket}" is
defined in terms of the two particles Wannier basis in the form:%
\begin{equation}
\left\vert k_{1},k_{2}\right\rangle =\frac{1}{2\pi }\sum\limits_{s_{1},s_{2}%
\in \mathcal{Z}}e^{ik_{1}s_{1}}e^{ik_{2}s_{2}}\left\vert
s_{1},s_{2}\right\rangle ,  \label{fourier1}
\end{equation}%
here $\mathcal{Z}$ is the set of integer numbers. Thus Eq.(\ref{QME}) for
two particles can be written as:%
\begin{equation}
\frac{d}{dt}\!\left\langle k_{1},\!k_{2}\left\vert \rho \left( t\right)
\right\vert k_{1}^{\prime },\!k_{2}^{\prime }\right\rangle \!= \!\mathcal{F}%
(k_{1},\!k_{1}^{\prime },\!k_{2},\!k_{2}^{\prime })\!\!\left\langle
k_{1},k_{2}\left\vert \rho \left( t\right) \right\vert k_{1}^{\prime
},k_{2}^{\prime }\right\rangle .  \label{F2}
\end{equation}%
Using the IC (\ref{IC1}) and the braket (\ref{fourier1}) the solution in
Fourier basis is 
\begin{equation}
\left\langle k_{1},k_{2}\left\vert \rho \left( t\right) \right\vert
k_{1}^{\prime },k_{2}^{\prime }\right\rangle =e^{\mathcal{F}%
(k_{1},k_{1}^{\prime },k_{2},k_{2}^{\prime })t},  \label{solK}
\end{equation}%
where%
\begin{eqnarray}
\mathcal{F}(k_{1},k_{1}^{\prime },k_{2},k_{2}^{\prime }) &\equiv &\left[ 
\mathcal{F}^{(1)}(k_{1},k_{1}^{\prime })+\mathcal{F}^{(1)}(k_{2},k_{2}^{%
\prime })\right]  \nonumber \\
&+&2D[\mathbf{C}\left( k_{1}\!,k_{2}^{\prime }\right) +\mathbf{C}\left(
k_{2}\!,k_{1}^{\prime }\right)  \nonumber \\
&-&\mathbf{C}\left( k_{1}\!,k_{2}\right) -\mathbf{C}\left( k_{1}^{\prime
}\!,k_{2}^{\prime }\right) ].  \label{F22}
\end{eqnarray}%
We note that%
\begin{equation}
\mathcal{F}^{(1)}(k_{i},k_{i}^{\prime })\equiv \left[ \!\frac{-i}{\hbar }%
\!\left( \mathcal{E}_{k_{i}}\!-\!\mathcal{E}_{k_{i}^{\prime }}\right)
\!+\!2D\!\left( \mathbf{C}\left( k_{i}\!,k_{i}^{\prime }\right) -\!1\right)
\!\right] ,  \label{F1}
\end{equation}%
is the one-particle infinitesimal generator in the Fourier representation
where 
\begin{equation}
\mathcal{E}_{k_{i}}\equiv \Omega \left\{ 1-\cos k_{i}\right\} ,  \label{F11}
\end{equation}%
that is, the eigenenergy of the free particle labeled "$i$" in the lattice 
\cite{mm}. The function%
\begin{equation}
\mathbf{C}\left( k\!_{1},k_{2}\right) \equiv \cos \left(
k_{1}-\!k_{2}\right) \!,  \label{F111}
\end{equation}%
takes into account the interaction induced between the particles. Therefore
the second term in (\ref{F22}) represents the interaction between particles
mediated by the thermal bath. In order to get insight into the mathematical
meaning of this interaction we can solve a \textit{pseudo} infinitesimal
generator considering \textit{only} the interaction term: $2D[\mathbf{C}%
\left( k_{1}\!,k_{2}^{\prime }\right) +\mathbf{C}\left(
k_{2}\!,k_{1}^{\prime }\right) -\mathbf{C}\left( k_{1}\!,k_{2}\right) -%
\mathbf{C}\left( k_{1}^{\prime }\!,k_{2}^{\prime }\right) ]$, that comes
from (\ref{part-bath}). In this case the solution will be normalized, but
the eigenvalues are not necessarily positive. This contribution in the full
infinitesimal generator (\ref{F22}) is the responsible of cross-terms
producing coherence between the two-particles, see appendix B.

If we solve (\ref{F2}) with $D=0$ the solution will represent two \textit{%
free tight-binding} particles in the lattice (i.e., a unitary evolution). If
we solve (\ref{F2}) with $\Omega =0$ and neglect the mentioned interaction,
this case will represent a classical problem (two RWs). In addition, from (%
\ref{F2}) the result: $\langle k_{1},k_{2}|d\rho (t)/dt|k_{1},k_{2}\rangle
=0 $ says that the diagonal Fourier elements are constant in time (i.e., a
momentum-like conservation law). In the section V using the Wigner function,
we will comment this statement.

The elements of $\rho (t)$ can be calculated on the Wannier basis, then we
can write an analytical formula for $\rho (t)$ in the real lattice: $\langle
s_{1},s_{2}|\rho (t)|s_{1,}^{\prime }s_{2}^{\prime }\rangle $. \bigskip
Using 
\[
\arrowvert s_{1},s_{2}\rangle =\frac{1}{2\pi }\int_{-\pi }^{\pi }\int_{-\pi
}^{\pi }dk_{1}dk_{2}\ e^{-ik_{1}s_{1}}e^{-ik_{2}s_{2}}\arrowvert %
k_{1},k_{2}\rangle , 
\]%
in the general solution of the QME (\ref{solK}), we get $\!\rho (t)$ in
Wannier's basis

\begin{eqnarray}
\left\langle s_{1},s_{2}\left\vert \!\rho (t)\!\right\vert s_{1}^{\prime
},s_{2}^{\prime }\right\rangle \! &=& \!\frac{1}{\left( 2\pi \right) ^{4}}\!
\int\int\int\int \limits_{-\pi }^{+\pi }\prod\limits_{i=1}^{2}dk_{i}\!\!\
dk_{i}^{\prime }\!\!\ e^{ik_{1}s_{1}-ik_{1}^{\prime }s_{1}^{\prime }} 
\nonumber \\
&\times &e^{ik_{2}s_{2}-ik_{2}^{\prime }s_{2}^{\prime }}e^{\mathcal{F}\left(
k_{1},k_{1}^{\prime },k_{2},k_{2}^{\prime }\right) t},  \nonumber \\
&&\quad \{s_{j},s_{l}^{\prime }\}\in \emph{Z}.  \label{roL1L2}
\end{eqnarray}%
These integrals can be done analytically considering Bessel's properties:%
\[
e^{iz\cos \theta }=\sum_{n=-\infty }^{\infty }i^{n}J_{n}(z)e^{in\theta
};\;\;e^{z\cos \theta }=\sum_{n=-\infty }^{\infty }I_{n}(z)e^{in\theta }, 
\]%
where $J_{n}$ and $I_{n}$ are Bessel's functions of integer order $n\in 
\mathcal{Z}$. These functions satisfies that \cite{evangelidis}%
\[
J_{-n}(x)=(-1)^{n}J_{n}(x),\ J_{n}(-x)=(-1)^{n}J_{n}(x), 
\]%
and 
\[
I_{-n}(x)=I_{n}(x),\ I_{n}(-x)=(-1)^{n}I_{n}(x). 
\]%
Then, we can write finally a closed expression for $\langle s_{1},s_{2}|\rho
(t)|s_{1,}^{\prime }s_{2}^{\prime }\rangle $.

To simplify the notation we use $t_{\Omega }\equiv \frac{\Omega t}{\hbar }%
,t_{D}\equiv 2Dt$ whenever it is necessary, then

\begin{eqnarray}
\langle s_{1},s_{2}|\rho (t)|s_{1}^{\prime },s_{2}^{\prime }\rangle \!
&=&\!i^{(s_{1}-s_{1}^{\prime }+s_{2}-s_{2}^{\prime
})}e^{-2t_{D}}\!\!\!\!\!\!\!\!\!\!\!\!\!\!\!\!\!\!\!\!\sum_{%
\{n_{1},n_{2},n_{3},n_{4},n_{5},n_{6}\}\in \mathcal{Z}}\!\!\!\!\!\!\!\!\!\!%
\!\!\!\!\!\!(-1)^{n_{4}+n_{5}}\   \nonumber \\
&\times &J_{s_{1}+n_{1}+n_{2}+n_{5}}\!\!\left( t_{\Omega }\right) \!\!\
J_{s_{1}^{\prime }+n_{1}+n_{3}+n_{4}}\!\!\left( t_{\Omega }\right)  
\nonumber \\
&\times &\!\!J_{s_{2}+n_{3}-n_{5}+n_{6}}\!\!\left( t_{\Omega }\right) \!\!\
J_{s_{2}^{\prime }+n_{2}-n_{4}+n_{6}}\!\!\left( t_{\Omega }\right) \!\! 
\nonumber \\
&\times &\prod\limits_{n_{i}=1}^{6}I_{n_{i}}\!\left( t_{D}\right) ,\
\{s_{j},s_{l}^{\prime }\}\in \mathcal{Z}.  \label{rhog}
\end{eqnarray}%
This solution is symmetric under the exchange of particles \footnote{%
To prove the invariance under the exchange of particles: $\left\{
s_{1}\leftrightarrow s_{2},s_{1}^{\prime }\leftrightarrow s_{2}^{\prime
}\right\} $, note that in Eq.(\ref{rhog}) $n_{j}$ are mute variables
therefore we can use the change of variables $n_{1}\leftrightarrow
n_{6},n_{2}\leftrightarrow n_{3}$ and finally $n_{4}\leftrightarrow
-n_{4},n_{5}\leftrightarrow -n_{5}$ to check this symmetry.} (i.e.,
preserving the symmetry of the IC). Of course, $\rho (t)$ is Hermitian,
positive definite and satisfies normalization in the lattice, that is:%
\begin{eqnarray}
\langle s_{1},s_{2}|\rho (t)|s_{1}^{\prime },s_{2}^{\prime }\rangle \!
&=&\!\langle s_{1}^{\prime },s_{2}^{\prime }|\rho ^{\ast
}(t)|s_{1},s_{2}\rangle ,\ \{s_{1},s_{2}\}\in \mathcal{Z}  \nonumber
\label{Ex-SOL2} \\
\mbox{Tr}\lbrack \rho (t)] &=&\!\!\!\!\sum_{\{s_{1},s_{2}\}\in \mathcal{Z}%
}\langle s_{1},s_{2}|\rho (t)|s_{1},s_{2}\rangle =1,\forall t,  \nonumber \\
&&
\end{eqnarray}%
the last line can be proved using Bessel's properties:%
\begin{eqnarray}
\sum_{n=-\infty }^{\infty }I_{n+m}(x)I_{n}(x) &=&I_{m}(2x)  \label{bessel1}
\\
\sum_{n=-\infty }^{\infty }(-1)^{n}I_{n+m}(x)I_{n}(x) &=&\delta _{m,0}.
\label{bessel2}
\end{eqnarray}%
The fact that $\rho (t)$ is positive definite, for all $t\geq 0$, follows
from the structural theorem when it is applied to our bonafide semigroup (%
\ref{QME}) \cite{Alicki}.

The probability of finding one particle in site $s_{1}$ and another in $%
s_{2} $ is given by probability profile:%
\[
P_{s_{1},s_{2}}(t)\equiv \langle s_{1},s_{2}|\rho (t)|s_{1},s_{2}\rangle , 
\]%
and shows for the present IC (\ref{IC1}) the expected reflection symmetry in
the plane: $s_{1}-s_{2}=0$.

Equation (\ref{rhog}) contains all the information concerning the
bath-induced correlations (off-diagonal elements). Note that $H_{\mathcal{S}%
} $ in (\ref{TH-H}) represents free distinguishable particles, the
particle-particle correlations are bath induced from (\ref{part-bath}).

In the case $D=0$, i.e., a closed system without dissipation, we recover the
density matrix for two QWs (the tight-binding solution):%
\[
\langle s_{1},s_{2}|\rho (t)|s_{1}^{\prime },s_{2}^{\prime }\rangle
_{D=0}=\prod_{j=1}^{j=2}i^{(s_{j}-s_{j}^{\prime })}J_{s_{j}}\!\left(
t_{\Omega }\right) \!J_{s_{j}^{\prime }}\!\left( t_{\Omega }\right) . 
\]%
This means that for a localized IC and in a closed system, $\rho (t)$ can be
written for any time $t>0$ as direct product of two independent particles,
i.e., $\rho (t)=\rho _{1}(t)\otimes \rho _{2}(t)$ .

Interestingly, in the case $D\rightarrow \infty $ (or $\Omega \rightarrow 0$%
) the classical regimen (two RWs) is not recovered \cite{Reichl,libro}
because $\mathcal{B}$ has created quantum coherence between them. Thus, in
the case $D\gg \Omega /\hbar $ the solution $\rho (t)$ is not the direct
product of two particles $\rho (t)\neq \rho _{1}(t)\otimes \rho _{2}(t)$,

\begin{eqnarray}  \label{rhoD}
\langle s_{1},s_{2}|\rho (t)|s_{1}^{\prime },s_{2}^{\prime }\rangle _{\Omega
=0}\! &=&\! (-1)^{(s_{1}+s_{1}^{\prime })}\delta _{s_{1}+s_{2},s_{1}^{\prime
}+s_{2}^{\prime }}\ e^{-2t_{D}}  \nonumber \\
&\times &\!\!\!\!\!\!\!\!\sum_{\{n_{1},n_{2},n_{3}\}\in \mathcal{Z}}\!\!\!\!
\!\!(-1)^{n_{1}+n_{2}}\!\ I_{s_{1}+n_{1}+n_{3}}\!\left( t_{D}\right) 
\nonumber \\
&\times &\!\!I_{-s_{2}+n_{2}+n_{3}}\!\left( t_{D}\right) \!\
I_{n_{1}}\!\left( t_{D}\right) \!\ I_{n_{2}}\!\left(t_{D}\right)  \nonumber
\\
&\times &\!\! I_{s_{1}-s_{1}^{\prime }+n_{1}+n_{2}+n_{3}}\!\left(
t_{D}\right) \!\ I_{n_{3}}\!\left( t_{D}\right) .  \nonumber \\
\end{eqnarray}%
Then, showing a complex pattern structure in terms of convolutions of
classical profiles.

From Eq.(\ref{rhoD}) we note that when $D\gg 1$ we get the probability
profile:%
\begin{eqnarray*}
P_{s_{1},s_{2}}(D\gg 1)\! &\neq &\!P_{s_{1}}(D\gg 1)\!\times
\!P_{s_{2}}(D\gg 1)=e^{-4Dt}\  \\
&\times &\!\!I_{s_{1}}\!\left( 2Dt\right) \!I_{s_{2}}\!\left( 2Dt\right) ,
\end{eqnarray*}%
here $P_{s_{j}}$ is the classical probability for each particle $j=1,2$ in
the sites $s_{1}$ and $s_{2}$ respectively. This result explicitly shows
that in the asymptotic regime $t\rightarrow \infty $ the classical profile
is not obtained. Thus the profile of probability for two DQW will not be a
Gaussian distribution. As we have commented before this is so because of
off-diagonal elements in $\rho (t)$ have been generated by the evolution of
the QME (quantum coherence bath induced).

We note that due to the presence of the bath there is a competition between
dissipation and building up coherence. This issue will be analyzed using
different measures in the next subsections.

\subsection{Change of basis for $\protect\rho \left( t\right) $ for two DWQs}

An outstanding conclusion can be observed by introducing a change of basis $%
U_{a}^{\dagger }\rho (t)U_{a}$ in the representation of the two-particle
density matrix. Which in fact is a function of parameters $(D,\Omega )$ and
the time $t$, see Eq.(\ref{rhog}).

Among several possibilities, and in order to calculate eigenvalues rather
than the eigenvectors, here we define a (dynamic) change of basis using a
time-dependent unitary transformation $U_{a}$ characterized by the elements%
\begin{equation}
\langle s_{1},s_{2}\arrowvert U_{a}\arrowvert s_{1}^{\prime },s_{2}^{\prime
}\rangle =i^{\left( s_{1}+s_{2}+s_{1}^{\prime }+s_{2}^{\prime }\right)
}J_{s_{1}-s_{1}^{\prime }}(\!\frac{\Omega t}{\hbar })J_{s_{2}-s_{2}^{\prime
}}(\!\frac{\Omega t}{\hbar }),  \label{A1}
\end{equation}%
with 
\begin{equation}
\langle s_{1},s_{2}\arrowvert U_{a}U_{a}^{\dagger }\arrowvert s_{1}^{\prime
},s_{2}^{\prime }\rangle =\delta _{s_{1},s_{1}^{\prime }}\delta
_{s_{2},s_{2}^{\prime }}.
\end{equation}%
Then, considering as before the IC $\rho (t=0)=\arrowvert0,0\rangle \langle
0,0\arrowvert$, it can be proved that%
\begin{eqnarray}
\langle s_{1},\!s_{2}\arrowvert U_{a}^{\dagger }\rho (t)U_{a}\arrowvert %
s_{1}^{\prime },\!s_{2}^{\prime }\rangle \! &=&\!\!\!\sum_{s_{1}^{^{\prime
\prime }},s_{2}^{^{\prime \prime }}}\sum_{s_{1}^{^{\prime \prime \prime
}},s_{2}^{^{\prime \prime \prime }}}\langle s_{1},s_{2}\arrowvert %
U_{a}^{\dagger }\arrowvert s_{1}^{\prime \prime },s_{2}^{\prime \prime
}\rangle  \nonumber  \label{ARA} \\
&\times &\langle s_{1}^{\prime \prime },s_{2}^{\prime \prime }\arrowvert\rho
(t)\arrowvert s_{1}^{\prime \prime \prime },s_{2}^{\prime \prime \prime
}\rangle  \nonumber \\
&\times &\langle s_{1}^{\prime \prime \prime },s_{2}^{\prime \prime \prime }%
\arrowvert U_{a}\arrowvert s_{1}^{\prime },s_{2}^{\prime }\rangle  \nonumber
\\
&&  \nonumber \\
&=&\delta _{s_{1}+s_{2},s_{1}^{\prime }+s_{2}^{\prime }}\ \left( -1\right)
^{s_{2}^{\prime }-s_{2}}\ e^{-2t_{D}}  \nonumber \\
&\times &\sum_{n_{1},n_{2},n_{5}}\left( -1\right) ^{n_{2}+n_{5}}\
I_{n_{2}}\!(t_{D}\!)  \nonumber \\
&\times
&\!\!I_{n_{3}}\!(t_{D}\!)I_{n_{5}}\!(t_{D}\!)I_{n_{2}+n_{5}+s_{1}}\!(t_{D}\!)
\nonumber \\
&\times
&\!\!I_{n_{3}+n_{5}-s_{2}}\!(t_{D}\!)I_{n_{2}+n_{5}+n_{3}+s_{2}^{\prime
}-s_{2}}\!(t_{D}\!),  \nonumber \\
&&
\end{eqnarray}%
in the last equality we have used Bessel's identity $\sum_{s}J_{s+m}(%
\!x)J_{s+p}(\!x)=\delta _{m,p}$. Comparing (\ref{ARA}) with (\ref{rhoD}) we
can conclude that in this new representation 
\begin{equation}
\tilde{\rho}(t)\equiv U_{a}^{\dagger }\rho (t)U_{a}=\rho (\Omega =0,D,t),
\label{A2}
\end{equation}%
i.e., $\tilde{\rho}(t)$ does not depend on the \textit{tigth-binding} energy 
$\Omega $. Then, it is simple to see that the new $\tilde{\rho}(t)$ is
Hermitian and equal to the highly dissipative case, and of course normalized
to one. In addition, it can be proved that purity and entropy are invariant
under this unitary transformation \cite{1057}.

This is an interesting result because allows us to study many quantum
properties using a simplest expression for the density matrix instead of
carrying on the analysis with the two parameters $(D,\Omega )$. Properties
as purity, entropy, etc., can straightforwardly be understood in this new
representation. We note that using $U_{a}$, eigenvectors also will change,
then a \textit{partial} trace would be affected by this map. But in the
present paper, we do not calculate any partial trace using $U_{a}$.

\subsection{Reduced Density Matrix for One Particle.}

At this point it is interesting to calculate from the full expression Eq.(%
\ref{rhog}), the reduced density matrix for one particle. This analysis will
help to understand the model and ultimately the induced coherence in the
system. To do this we trace out over the degrees of freedom of one particle,
say $j=2$. Then, the reduced density matrix for one particle is obtained as%
\begin{equation}
\langle s_{1}|\rho ^{(1)}(t)|s_{1}^{\prime }\rangle \!=\!\langle s_{1}%
\arrowvert\mbox{Tr}_{2}[\rho (t)]\arrowvert s_{1}^{\prime }\rangle
\!=\!\sum_{s_{2}\in \mathcal{Z}}\langle s_{1},s_{2}|\rho (t)|s_{1}^{\prime
},s_{2}\rangle .  \label{Ro1}
\end{equation}%
Alternatively, note that using the Fourier expression Eq.(\ref{roL1L2}) it
is straightforward to find that%
\[
\langle s_{1}|\rho ^{(1)}(t)|s_{1}^{\prime }\rangle \!=\!\!\left( 2\pi
\right) ^{-2}\!\!\!\int \!\!\!\!\int \!\!dk_{1}\!\!\ dk_{1}^{\prime }\!\!\
e^{ik_{1}s_{1}-ik_{1}^{\prime }s_{1}^{\prime }}e^{\mathcal{F}^{(1)}\left(
k_{1},k_{1}^{\prime }\right) t}, 
\]%
with $\mathcal{F}^{(1)}(k_{1},k_{1}^{\prime })\!\!=\!\!\{\!\frac{-i}{\hbar }%
\!\left( \mathcal{E}_{k_{1}}\!-\!\mathcal{E}_{k_{1}^{\prime }}\right)
\!+\!2D\!\left( \cos \left( k_{1}\!-\!k_{1}^{\prime }\right) \!-\!1\right)
\!\}$, the one-particle infinite generator. Then, we arrive to the result%
\begin{equation}
\langle s_{1}|\rho ^{(1)}\!(t)|s_{1}^{\prime }\rangle
\!\!=\!i^{(s_{1}-s_{1}^{\prime })}e^{-t_{D}}\!\!\!\sum_{n\in \mathcal{Z}%
}\!\!J_{s_{1}+n}\!\left( t_{\Omega }\right) \!\!\ J_{s_{1}^{\prime
}+n}\!\left( t_{\Omega }\right) \!\!\ I_{n}\!\left( t_{D}\right) \!.
\label{1-part}
\end{equation}

This marginal solution corresponds to the one-particle density matrix with
IC $\rho (t=0)=\arrowvert0\rangle \langle 0\arrowvert$, and shows when $D\gg
\Omega /\hbar $ asymptotically the same behavior than a classical RW \cite%
{mm}. A result that is not entirely surprising because in the Hamiltonian (%
\ref{Ht}) each particle is originally non-interacting between them. We
remain that the correlations are built up between the particles as a result
of the interaction with $\mathcal{B}$. In the next section we will show that
the two-body profile does not behave as two classical random walks in any
asymptotic regime.

\subsection{Eigenvalues of $\protect\rho (t)$}

Here we calculate the eigenvalues of the density matrix for any fixed time.
In the case of one-particle the expression for the eigenvalues is
analytical. On the other hand, in the two-particles case, numerical
calculations can be done using our exact result written in terms of Bessel's
functions.

\subsubsection{The one-body reduced density matrix case}

Consider the solution of the one-particle reduced density matrix (\ref%
{1-part}). We want to find a new representation where $\rho ^{(1)}(t)\!$ is
diagonal for any fixed time $t$; that is, we want to find a unitary
transformation $U$ such that%
\[
U^{\dagger }\rho ^{(1)}U=\tilde{\rho}^{(1)}. 
\]%
Thus, using the solution (\ref{1-part}) we can explicitly show (where $%
t_{D}=2Dt,t_{\Omega }\equiv \Omega t/\hbar $) that%
\begin{eqnarray}
\left( U^{\dagger }\rho ^{(1)}U\right) _{nk} &=&\langle n|U^{\dagger }\rho
^{(1)}U|k\rangle =\sum_{r,q}\left( U^{\dagger }\right) _{nr}\rho
_{rq}^{(1)}U_{qk}  \nonumber \\
&=&\!e^{-t_{D}}\!\!\sum_{r,q}\left( U^{\dagger }\right) _{nr}U_{qk}\
i^{(r-q)}\!\!\sum_{\alpha \in \mathcal{Z}}\!J_{r+\alpha }\!\left( t_{\Omega
}\right)  \nonumber \\
&\times &J_{q+\alpha }\!\left( t_{\Omega }\right) \!\!\ I_{\alpha }\!\left(
t_{D}\right)  \nonumber \\
&=&e^{-t_{D}}\sum_{\alpha \in \mathcal{Z}}I_{\alpha }\!\left( t_{D}\right)
\sum_{r}i^{r}U_{rn}^{\ast }\ J_{r+\alpha }\!\left( t_{\Omega }\right) 
\nonumber \\
&\times &\sum_{q}i^{-q}U_{qk}\ J_{q+\alpha }\!\left( t_{\Omega }\right)
\!\!\ .  \label{auto1}
\end{eqnarray}%
Noting that $\sum_{n=-\infty }^{\infty }J_{r+n}(x)J_{r+\alpha }(x)=\delta
_{n,\alpha }$, we see from (\ref{auto1}) that if 
\begin{equation}
U_{qk}=i^{(q+k)}J_{q+k}\left( t_{\Omega }\right) ,  \label{auto0}
\end{equation}%
we get $U^{\dagger }U=\mathbf{1}$ (unitary transformation), then 
\begin{eqnarray}
\left( U^{\dagger }\rho ^{(1)}U\right) _{nk} &=&e^{-t_{D}}\sum_{\alpha \in 
\mathcal{Z}}I_{\alpha }\!\left( t_{D}\right) \ i^{k-n}\delta _{n,\alpha
}\delta _{k,\alpha }  \nonumber \\
&=&e^{-t_{D}}I_{n}\!\left( t_{D}\right) \ \delta _{n,k}  \label{auto2}
\end{eqnarray}%
This means that the eigenvalues and their normalized eigenvectors are 
\begin{eqnarray}
\lambda _{0} &=&e^{-t_{D}}I_{0}\!\left( t_{D}\right) \ \;\;\rightarrow
\;(\cdots ,i^{k}J_{k}\left( t_{\Omega }\right) ,\cdots )^{T},  \nonumber \\
\lambda _{\pm 1} &=&e^{-t_{D}}I_{\pm 1}\!\left( t_{D}\right) \ \rightarrow
\;(\cdots ,i^{\pm 1+k}J_{\pm 1+k}\left( t_{\Omega }\right) ,\cdots )^{T}, 
\nonumber \\
\lambda _{\pm 2} &=&e^{-t_{D}}I_{\pm 2}\!\left( t_{D}\right) \rightarrow
\;(\cdots ,i^{\pm 2+k}J_{\pm 2+k}\left( t_{\Omega }\right) ,\cdots )^{T}, 
\nonumber \\
&&\mbox{with}\;\;k\in 0,\pm 1,\pm 2,\cdots \;\;\;\mbox{etc.}  \label{eigen}
\end{eqnarray}%
We note that $\lambda _{n}\in \mathcal{R}_{e}\ (\forall n\in 0,\pm 1,\pm
2,\cdots )$ and that they are bounded in the interval $1\geq \lambda
_{n}\geq 0.$

\begin{figure}[t]
\centering
\includegraphics[width=0.5 \columnwidth,clip]{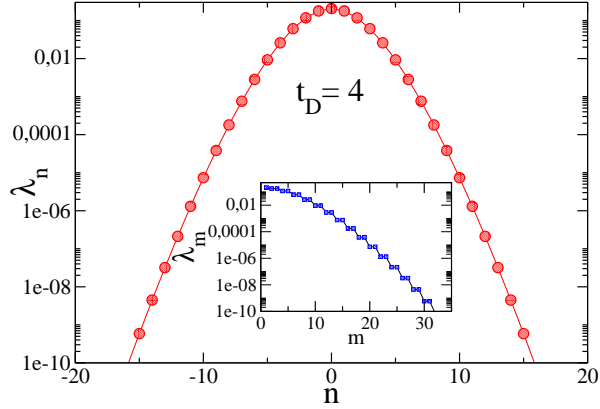}
\caption{(Color online) Analytical eigenvalues (in log scale) of the
one-body reduced density matrix. The inset shows the numerical calculation
ordered from the largest eigenvalue.}
\label{fig1}
\end{figure}

In figure \ref{fig1} we show the eigenvalues $\lambda _{n}$ of the $1$-body
reduced density matrix $\rho ^{(1)}\left( t\right) $ for $t_{D}=2Dt=4$. The
inset in this figure shows the plot of the numerical calculation ordered
from the largest one. We can see that the eigenvalues are degenerate in
pairs, except for the first eigenvalue. This fact can be understood as
follows: by means of the unitary transformation (\ref{auto0}) we can write $%
\rho \left( t\right) \ $in a diagonal form using the eigenvalues (\ref{eigen}%
) 
\begin{eqnarray*}
\tilde{\rho} &=&\sum_{n=-\infty }^{+\infty }\lambda _{n}\arrowvert n\rangle
\langle n\arrowvert \\
&=&\sum_{n=-\infty }^{-1}\lambda _{n}\arrowvert n\rangle \langle n\arrowvert%
+\sum_{n=1}^{+\infty }\lambda _{n}\arrowvert n\rangle \langle n\arrowvert%
+\lambda _{0}\arrowvert0\rangle \langle 0\arrowvert \\
&=&\sum_{n=1}^{+\infty }\lambda _{n}\arrowvert n\rangle \langle n\arrowvert%
+\sum_{n=1}^{+\infty }\lambda _{-n}\arrowvert-n\rangle \langle -n\arrowvert%
+\lambda _{0}\arrowvert0\rangle \langle 0\arrowvert,
\end{eqnarray*}%
then, we conclude that $\tilde{\rho}$, with the IC (\ref{IC1}), is invariant
under reflection symmetry $\arrowvert n\rangle \rightarrow \arrowvert%
-n\rangle $, and so the eigenvalues $\lambda _{n}$ are degenerate in pairs $%
\lambda _{n}=\lambda _{-n}$ except $\lambda _{0}$ as it is shown in figure %
\ref{fig1}. 

Also from (\ref{auto0}) we see that the density matrix is normalized: 
\begin{equation}
\mbox{Tr}\left[ \tilde{\rho}\right] =\sum_{n\in \mathcal{Z}}\lambda
_{n}=\sum_{n\in \mathcal{Z}}e^{-t_{D}}I_{n}\!\left( t_{D}\right) =1.
\label{trasa1}
\end{equation}

Using this new basis we can write an analytical expression for the
one-particle purity and entropy (they are invariant under unitary
transformations \cite{1057}). In this basis%
\begin{eqnarray}
\left( \tilde{\rho}^{(1)}\right) ^{2}\! &=&\!\left( \sum_{n\in \mathcal{Z}%
}\lambda _{n}\arrowvert n\rangle \langle n\arrowvert\right)
^{2}\!=\!\sum_{n,m\in \mathcal{Z}}\lambda _{n}\lambda _{m}\arrowvert %
n\rangle \langle n\arrowvert m\rangle \langle m\arrowvert  \nonumber \\
&=&\sum_{n\in \mathcal{Z}}\lambda _{n}^{2}\arrowvert n\rangle \langle n%
\arrowvert,  \label{pury1}
\end{eqnarray}%
therefore, using (\ref{bessel1}), the purity is%
\begin{eqnarray}
\mathcal{P}_{Q}^{(1)}\left( t\right) =\mbox{Tr}\left[ \left( \tilde{\rho}%
^{(1)}\right) ^{2}\right] &=&\sum_{n\in \mathcal{Z}}e^{-2t_{D}}\ \left[
I_{n}\!\left( t_{D}\right) \right] ^{2}  \nonumber \\
&=&e^{-2t_{D}}I_{0}\!\left( 2t_{D}\right) .  \label{purity}
\end{eqnarray}

In an analog way the entropy can also be calculated analytically 
\begin{eqnarray}
S^{(1)} \!\!&=&\!\!-\mbox{Tr}\left[ \tilde{\rho}^{(1)}\ln \tilde{\rho}^{(1)}%
\right]  \nonumber \\
\!\! &=&\!\!-\!\!\sum_{n\in \mathcal{Z}}\lambda _{n}\ln \lambda
_{n}\!\!=\!-\!\!\sum_{n\in \mathcal{Z}}e^{-t_{D}}I_{n}\!\left( t_{D}\right)
\ln \left( e^{-t_{D}}I_{n}\!\left( t_{D}\right) \right)  \nonumber \\
&=&-\sum_{n\in \mathcal{Z}}e^{-t_{D}}I_{n}\!\left( t_{D}\right) \left[
-t_{D}+\ln \left( I_{n}\!\left( t_{D}\right) \right) \right]  \nonumber \\
&=&t_{D}-e^{-t_{D}}\sum_{n\in \mathcal{Z}}I_{n}\!\left( t_{D}\right) \ln
\left( I_{n}\!\left( t_{D}\right) \right) ,  \label{s1}
\end{eqnarray}%
where we have used (\ref{eigen}).

To end this section we comment that (\ref{s1}) agrees with numerical
calculations presented in \cite{mm} ($S^{(1)}$ is linear for $t_{D}<<1$). In
particular when $D=0$ and noting that $I_{n}\!\left( 0\right) =\delta _{n,0}$
we get that $S^{(1)}=0$, and in general for $D\neq 0$ we get $S^{(1)}\left(
t\right) >0,$ $\forall t>0.$

\subsubsection{The two-body density matrix case}

\begin{figure}[t]
\centering
\includegraphics[width=0.5 \columnwidth,clip]{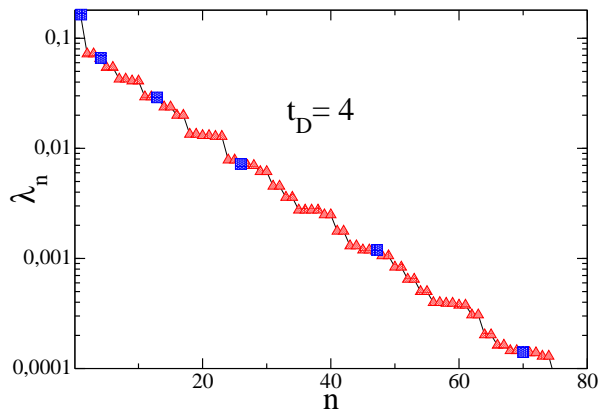}
\caption{(Color online) Eigenvalues of the two-body (reduced) density
matrix. The squares represent the non-degenerated eigenvalues. }
\label{fig2}
\end{figure}

Consider the two-particle density matrix (\ref{rhog}). Here we want to find
the eigenvalues of $\rho (t)\!$ for any fixed time $t$. Using the unitary
transformation $U_{a}$ presented in (\ref{A1}) the two-particle density
matrix can be written in the form $\tilde{\rho}(t)\equiv \rho (\Omega
=0,D,t) $, as was proved in (\ref{A2}). Unfortunately, we were not able to
find an analytical expression for these eigenvalues in the case of two
particles, but its analysis can be done numerically from $\tilde{\rho}(t)$.
In figure \ref{fig2} we show the eigenvalues of the two-body density matrix $%
\tilde{\rho}(t)$ for $t_{D}=2Dt=4$. Comparing this figure with the inset of
figure \ref{fig1}, it can be realized the complex structure for the two-body
eigenvalues. It can be seen that there are degenerated eigenvalues in pairs,
and also non-degenerated values (see figure \ref{fig2}). In order to
understand these degererancy we have carried out a numerical analysis of
eigenvectors. Then, we can conclude that the symmetry that is behind the
two-particles eigenvalue is also the refection symmetry of the two-particle
eigenvector. That is, consider the Wannier ket $\arrowvert %
s_{i},s_{j}\rangle $ and the symmetry $s_{i}+s_{j}=n$, a two-particle
eigenvector can be written in the form: 
\[
\arrowvert c_{1},c_{2}\rangle =b\arrowvert s_{1},s_{2}\rangle +b^{\prime }%
\arrowvert s_{1}^{\prime },s_{2}^{\prime }\rangle +\cdots , 
\]%
such that 
\[
s_{1}+s_{2}=n,\ s_{1}^{\prime }+s_{2}^{\prime }=n,\cdots ,\ \mbox{with }%
n=1,2,\cdots , 
\]%
and its corresponding eigenvalue is degenerated with the case$\
m=-1,-2,\cdots $. The non-degenerated eigenvalues correspond to eigenvectors
in the subspace with $s_{i}+s_{j}=0$. It means that $\tilde{\rho}\left(
t\right) $ is diagonalizable in blocks with $s_{1}+s_{2}=s_{1}^{\prime
}+s_{2}^{\prime }=\cdots $. 

\section{Quantum coherence}

To study the coherence we will use standard measures to characterize the
process. Among the different measures, here we present the ones that allow
us analytical results: Profile probability,\ von Neuman entropy, Purity,
Quantum coherence, and Spatiotemporal dispersion. Non-classical correlations
and the negativity of the Wigner function will we presented in separated
sections.

\subsection{Purity}

To see the influence of $\mathcal{B}$ to build up coherence between the
particles, we calculate the quantum purity $\mathcal{P}_{Q}(t)\equiv %
\mbox{Tr}\lbrack \rho (t)^{2}]$. Linear entropy or impurity of the system $%
\mathcal{S}_{L}=1-\mathcal{P}_{Q}$ is a lower approximation to the quantum
von Neuman entropy $-\mbox{Tr}\lbrack \rho \ln \rho ]$, then if system
remains pure ($\mathcal{P}_{Q}=1$) or mixed ($\mathcal{P}_{Q}<1$). In our
case we can study analytically this two-body quantity in the course of time:

\begin{eqnarray}
\mathcal{P}_{Q}^{(2)}(t)\!=\!\!\!\!\!\!\! &&e^{-4t_{D}}\sum_{m\in \mathcal{Z}%
}I_{m}\left( 2t_{D}\right) \sum_{\{\alpha ,\beta \}\in \mathcal{Z}%
}(-1)^{\alpha +\beta }I_{\alpha }\left( 2t_{D}\right)  \nonumber \\
&\times &\! I_{\beta }\left( 2t_{D}\right) I_{\alpha +m}\left( 2t_{D}\right)
I_{\beta +m}\left( 2t_{D}\right) I_{\alpha +\beta +m}\left( 2t_{D}\right) . 
\nonumber \\
&&  \label{P2}
\end{eqnarray}

It can be seen for $D=0$ (without dissipation) that purity $\mathcal{P}%
_{Q}^{(2)}(t)$ takes the value one for all time. But for the case $D\neq 0$
the purity is lower than one and decreases in time. For $D\neq 0$ the purity
is different from the purity for \textit{two-particles} with independent
quantum bath, i.e., 
\[
\mathcal{P}_{Q}^{(2)}(t)\neq \mathcal{P}_{Q}^{(1)}(t)\mathcal{P}%
_{Q}^{(1)}(t), 
\]%
here $\mathcal{P}_{Q}^{(1)}(t)$ is the corresponding one-particle purity
with independent bath $\mathcal{P}_{Q}^{(1)}(t)=e^{-2t_{D}}I_{0}\left(
2t_{D}\right) $ \cite{mm}. Therefore a common quantum bath $\mathcal{B}$ has
produced a difference $\Delta \mathcal{P}_{Q}=\mathcal{P}_{Q}^{(2)}(t)-%
\mathcal{P}_{Q}^{(1)}(t)\mathcal{P}_{Q}^{(1)}(t)$ which shows the occurrence
of classical and non-classical correlations between particles.

\begin{figure}[t]
\centering
\includegraphics[width=0.5 \columnwidth,clip]{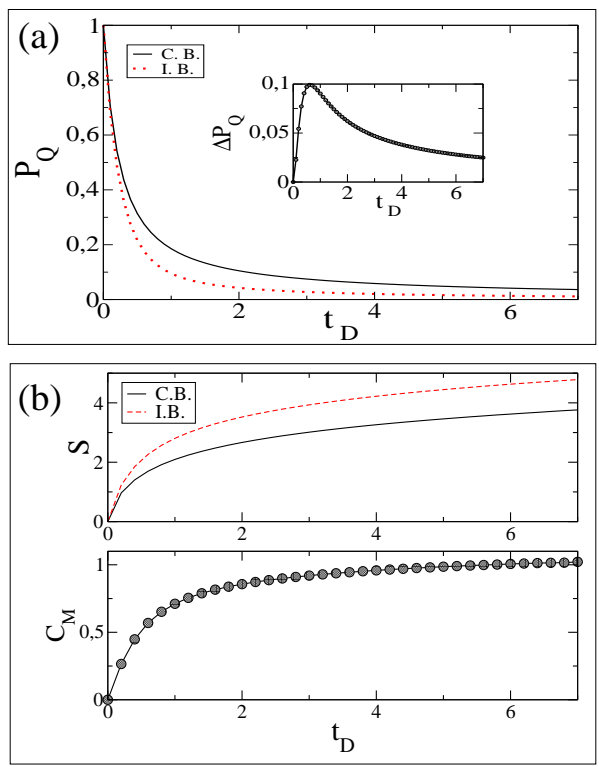}
\caption{(Color online) (a) Purity from a localized IC (\protect\ref{IC1})
as function of $t_{D}=2Dt$. Plots in straight lines are for the case with a
common bath, dashed lines are with independent baths. The inset shows the
difference of the Purity between the cases: two DQWs with a common $\mathcal{%
B}$ (C.B.), and two DQWs with independent baths (I.B.). \newline
(b) Entropy for a localized IC (\protect\ref{IC1}) as function of $t_{D}=2Dt$%
, for two different cases. Plots in straight line is for the case of two
particles with a common bath, dashed line is with independent baths. In the
bottom the QMI\ ($C_{M}$) is shown as a function of $t_{D}$. }
\label{fig3}
\end{figure}

The purity $\mathcal{P}_{Q}$ is related to the entropy $S(t)$ of the system.
In figure \ref{fig3}(a) we show that $\rho (t)$ for two particles with a 
\textit{common} bath has more purity than in the case of two particles with 
\textit{independent} baths. This fact can be thought as a measure of the
bath induced coherence between the particles. The inset in figure \ref{fig3}%
(a) shows the difference of the purity $\Delta \mathcal{P}_{Q}$ between the
two mentioned cases, showing that $\Delta \mathcal{P}_{Q}$ has in fact a
maximum of coherence and then decreases slowly. Therefore, for $t_{D}^{\max
}=2Dt^{\max }\sim 1$ there is a characteristic time-scale before the effect
of dissipation wipe out the coherence induced by the bath. This result will
be compared in the next sections with the coherence measured from the
off-diagonal elements of the two-body density matrix, and the negative
volume of the Wigner function. 

The figure \ref{fig3}(b) the entropy and the Quantum Mutual Information
(QMI) are also plotted as a function of $t_{D}$. On the top of this figure
the entropy is shown for two particles with common bath $S^{\left( 12\right)
}(t)$ and independent baths $2S^{\left( 1\right) }(t)$. In the bottom of
this figure, we plot $C_{M}(t)$ which quantify the QMI:%
\[
C_{M}(t)=S^{\left( 1\right) }(t)+S^{\left( 2\right) }(t)-S^{\left( 12\right)
}(t)=2S^{\left( 1\right) }(t)-S^{\left( 12\right) }(t). 
\]%
The QMI measures the total correlation (quantum and classical) in the
system. Then, we can conclude from the present analysis that at short times
when there is not too coherence in the system (initially particles are
uncorrelated) the QMI grows up fast, but as soon as particles acquire
bath-induced coherence between them, the QMI starts increasing slowly.

\subsection{Spatial Correlation Induced by the Bath.}

It is now convenient to define a new measure that quantifies the
spatiotemporal correlation function between two distinguishable DQWs. To do
this we define the function:%
\begin{equation}
\mathcal{C}_{(1,2)}=\left\langle \widehat{q}_{1}^{2}\widehat{q}%
_{2}^{2}\right\rangle -\left\langle \widehat{q}_{1}^{2}\right\rangle
\left\langle \widehat{q}_{2}^{2}\right\rangle ,  \label{SpaCorre}
\end{equation}%
where $\widehat{q}_{j}$ is the position operator for each particle $j=1,2$.
This operator is diagonal in the Wannier basis: 
\[
\widehat{q}_{j}|s_{1},s_{2}\rangle =s_{j}|s_{1},s_{2}\rangle ,j=1,2. 
\]%
Then $\mathcal{C}_{(1,2)}$ is zero for independent particles, as would be
for two RWs or Wiener processes. Otherwise any difference from zero ($%
\mathcal{C}_{(1,2)}\neq 0$) indicates a coherence of the two-body density
matrix (\ref{rhog}). We shall show that in fact the two (free) particles
build up\ spatiotemporal correlations as soon as the temperature of the bath
is larger than zero. We note that the quantity $\mathcal{C}_{(1,2)}$ is a
"semi-classical" measure because we are using distinguishable operators: $%
\widehat{q}_{j},\ j=1,2$. We now calculate $\mathcal{C}_{(1,2)}$ using Eq. (%
\ref{rhog}) and the IC (\ref{IC1}): 
\begin{equation}
\mathcal{C}_{(1,2)}=(2Dt)^{2}.  \label{SpaCorre1}
\end{equation}%
This result says that $\mathcal{B}$ induces coherence as soon as $t>0$. Here
it is interesting to remark that the scaling parameter is $D\equiv \Gamma
^{2}k_{B}T/\hbar $, ($T$ is the temperature of the bath $\mathcal{B}$) and
not any other combination of model parameters. We note that for a \textit{%
tight-binding} particle the second moment is $\left\langle \widehat{q}%
_{1}^{2}\right\rangle -\left\langle \widehat{q}_{1}\right\rangle ^{2}=\frac{1%
}{2}\left( \frac{\Omega t}{\hslash }\right) ^{2}+2Dt$, showing a ballistic
regime when there is not dissipation. To end this paragraph we want to
comment that in a classical \textit{correlated} RW model: $dx_{j}/dt=\xi
_{j}(t)$ with $\left\langle \xi _{j}(t)\xi _{i}(t^{\prime })\right\rangle
=c_{ij}\ \delta \left( t-t^{\prime }\right) $ and $c_{ij}\neq \delta _{ij}$,
the 4th moment would be $\left\langle x_{1}^{2}(t)x_{2}^{2}(t)\right\rangle
-\left\langle x_{1}^{2}(t)\right\rangle \left\langle
x_{2}^{2}(t)\right\rangle =4x_{1}(0)x_{2}(0)c_{12}\ t+\left( 2c_{12}t\right)
^{2}$ (here $x_{1}(0),x_{2}(0)$ are the initial conditions of the classical
particles). Comparing this last classical result with (\ref{SpaCorre1}) we
conclude that we cannot assert on the quantum nature of the process, other
measures will be needed to quantify the quantumness of the bath induced
correlations. This subject will be presented in next sections.

\subsection{Calculating Numerical Results for the Probability Profile}

\begin{figure}[t]
\centering
\includegraphics[width=0.4 \columnwidth,clip]{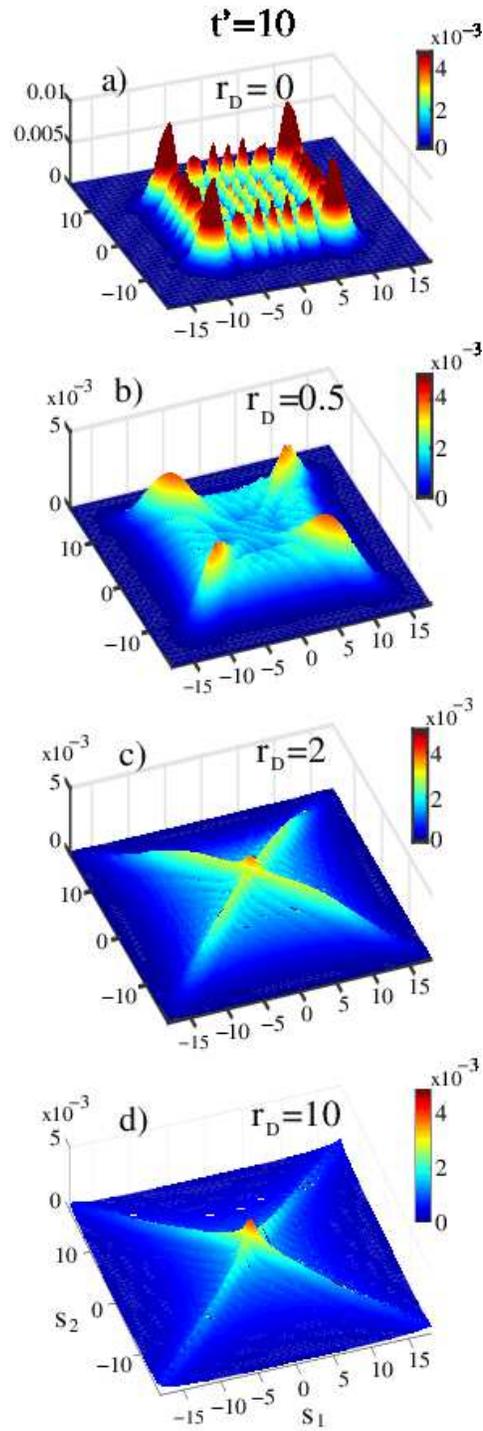}
\caption{(Color online) Probability profile $P_{s_{1},s_{2}}(t^{\prime })$
with $t^{\prime }=t_{\Omega }=10$, for an IC $\protect\rho (t=0)=\arrowvert%
\vec{0}\rangle \langle \vec{0}\arrowvert$ as a function of position of
particles $s_{1}$ and $s_{2}$, for (a) $r_{D}=0$, (b) $r_{D}=0.5$. The
interference pattern can be seen even in the presence of large dissipation:
(c) for $r_{D}=2$ and (d) $r_{D}=10$. Blue indicates, roughly, the value
zero while red the high value of probability.}
\label{fig-prob2DQW}
\end{figure}

It is convenient to define re-scaled parameters, which in fact help to
understand the complex dynamics of the two particles. Let $r_{D}$ be the
rate of characteristic energy scales in the system: $r_{D}\equiv \frac{2D}{%
\Omega /\hbar }$, and $t^{\prime }$ a dimensionless time $t^{\prime
}=t_{\Omega }$. In Fig.\ref{fig-prob2DQW} we show the probability of finding
particles at the site $s_{1}$ and $s_{2}$, i.e., $P_{s_{1},s_{2}}(t^{\prime
})=\langle s_{1},s_{2}\arrowvert\rho (t^{\prime })\arrowvert %
s_{1},s_{2}\rangle $ for four values of the dissipative parameter $r_{D}=0,$ 
$0.5,$ $2,$ $10$, see Eq.(\ref{rhog}). 
Fig.\ref{fig-prob2DQW}(a) corresponds to the case when the two free
particles do not interact with the bath ($D\equiv \Gamma ^{2}k_{B}T/\hbar =0$%
), here the evolution of particles are ballistic (not diffusive) and
characterized by Anderson' velocity: $V_{A}=\frac{1}{\sqrt{2}}\frac{\Omega }{%
\hslash }$ \cite{mm}, this is a pure quantum regime. When the temperature of
the bath $\mathcal{B}$ is different from zero the profile $%
P_{s_{1},s_{2}}(t^{\prime })$ is modified, appearing interference patterns
along of line $s_{1}=s_{2}$, and raising the value of the probability in the
direction $s_{1}=-s_{2}$ (conservation of total momentum), see Figs.\ref%
{fig-prob2DQW}(b),(c). In the case $r_{D}\gg 1$ (strong dissipation case)
the profile $P_{s_{1},s_{2}}(t^{\prime })$ shows a different interference
pattern signing the quantum nature of the behavior, see Fig.\ref%
{fig-prob2DQW}(d). This is in contrast to the case of two particles with
independent baths, in which case the profile would be a Gaussian
distribution. It is important to remark that for two DQWs with a common bath
the profile can never be represented as a Gaussian distribution for any
value of $r_{D}$ or $t^{\prime }$. This result says that $\mathcal{B}$
induces coherence between the particles while also producing dissipation.

\subsection{Cross terms of the two-body $\protect\rho (t)$.\textit{\ }}

\begin{figure}[t]
\centering
\includegraphics[width=0.5 \columnwidth,clip]{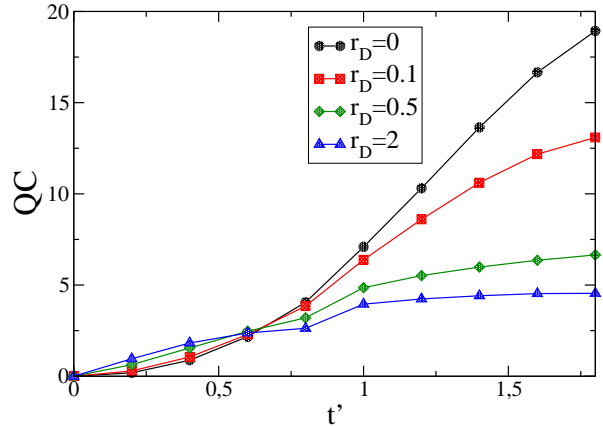}
\caption{(Color online) Quantum coherence (QC) for two particles with IC (%
\protect\ref{IC1}) as function of $t^{\prime }=t_{\Omega }\equiv \Omega
t/\hbar $. This function shows a crossover at $t^{\prime }\simeq 0.6$ as a
function of time. The plot shows $\mathcal{G}$ calculated for different
values of the dissipation parameter $r_{D}=\frac{2D}{\Omega /\hbar }.$}
\label{fig5}
\end{figure}

\begin{figure}[t]
\centering
\includegraphics[width=0.5 \columnwidth,clip]{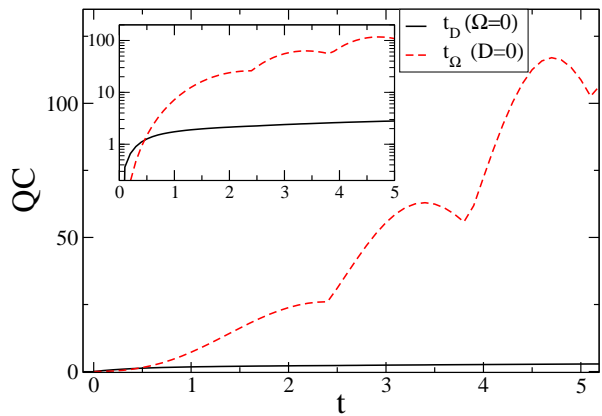}
\caption{(Color online) Quantum coherence (QC) for two particles with IC (%
\protect\ref{IC1}) for extrema cases: zero dissipation ($D=0$), and strong
dissipation ($\Omega =0$) as a function of $t^{\prime }=$ $t_{\Omega }\equiv
\Omega t/\hbar $. This function shows a notable structured behavior as a
function of time for the non-dissipative case. While in the strong
dissipative case the behavior is monotonous increasing, showing only a
suddenly increase at short times. In the inset, the function $\mathcal{G}$
is shown in logarithmic scale.}
\label{fig6}
\end{figure}

A measure to indirectly quantify the occurrence of correlations between the
particles can be evaluated by calculating the total coherence contribution
from the cross-terms of the density matrix. This object is defined as%
\[
\mathcal{G}=\sum_{\atop{}{{}{{\ (s_{1}\neq s_{1}^{\prime }) \newline
(s_{2}\neq s_{2}^{\prime })}}}}\left\vert \langle s_{1},s_{2}\arrowvert\rho
(t)\arrowvert s_{1}^{\prime },s_{2}^{\prime }\rangle \right\vert , 
\]%
and it has recently been used to quantify the quantum coherence \cite%
{Plenio2014,4746}. This measure is easy to compute than the relative
coherence entropy which need a diagonalization procedure \cite{bu}%
In figure \ref{fig5}, we show the Quantum Coherence (QC) $\mathcal{G}$ as a
function of $t^{\prime }=\frac{\Omega t}{\hbar }$ and for several values of
dissipation $r_{D}=0,$ $0.1,$ $0.5,$ $2$. As we will see with other measures
for $0<t^{\prime }<0.6\simeq \tau _{c}$ the QC is larger for the case $T\neq
0$ ($r_{D}\neq 0$) than with respect to the case $T=0$ ($r_{D}=0$). This
result also indirectly indicates that thermal bath has created correlations
between particles for $t^{\prime }<\tau _{c}$, and for times $t^{\prime
}>\tau _{c}$ will vanish for the presence of the bath dissipation. In order
to clarify the behavior of QC for $t^{\prime }>\tau _{c}$ (larger for $T=0$
than $T\neq 0$), the two extreme cases are shown in figure \ref{fig6}; that
is, the function $\mathcal{G}$ as function of $t_{\Omega }$ for the zero
dissipation case ($D=0$), and $\mathcal{G}$ as function of $t_{D}$ for the
strong dissipation case ($\Omega =0$). 

As can be seen from the exact result (\ref{rhog}), the solution of $\rho
\left( t\right) $ is a convolution of quantum and classical contributions.
The signature of the quantum character appears through J-Bessel's functions
which oscillates in time ($t_{\Omega }\equiv \frac{\Omega t}{\hbar }$),
while the classical functionality comes from I-Bessel's functions (note that
here the time appears through the quantity: $t_{D}\equiv 2Dt$). Therefore,
it is simple to realize that the nature of the oscillations in the case $D=0$
(see figure \ref{fig6}) comes from the temporal behaviors of the J-Bessel's
functions. On the contrary, in the case $\Omega =0$ the function $\mathcal{G}
$ is a smooth function of time (only depends on the I-Bessel's functions).

It is important to remark that the crossing-time that we have found
analyzing the QC is of the order of the scaling-time that we got from the
study of the purity (see figure \ref{fig3}).

Numerically, the long time behavior of the function $\mathcal{G}$ is very
hard to get (the solution $\rho \left( t\right) $ involves the product of
four J-Bessel's and six I-Bessel's functions in the lattice). Nevertheless,
it is not difficult to realize from (\ref{rhog}), that if $D\neq 0$ at long
time all elements of $\rho \left( t\right) $ go to zero (conserving
normalization see (\ref{Ex-SOL2})).

\section{On the Wigner phase-space representation.}

A novel point of view can be achieved if we introduce a quasi probability
distribution function (pdf) on the lattice \cite{Wootters}. A similar
representation was used for the case of one-particle $\rho (t)$ in reference 
\cite{mm}. The crucial point in the definition of a quasi-pdf is to assure
the completeness of the representation, a fact that can be proved from our
Wigner-like pdf.

Consider the quasi-pdf on the enlarged lattice of integers ($\mathcal{Z}$)
and semi-integers ($\mathcal{Z}_{2}$), and use the notation:%
\[
\vec{k}=(k_{1},k_{2}),\vec{x}=(x_{1},x_{2}),x_{j}\in (\mathcal{Z}\oplus 
\mathcal{Z}_{2}). 
\]%
We define%
\begin{eqnarray}
W\!(\vec{k},\vec{x},t)\! &=& \!(2\pi )^{-2}\!\!\!\!\!\!\!\!
\sum_{x_{1}^{\prime },x_{2}^{\prime }\in (\mathcal{Z}\oplus \mathcal{Z}%
_{2})} \!\!\!\!\!\!\!\!\!\langle x_{1}\!+\!x_{1}^{\prime
},x_{2}\!+\!x_{2}^{\prime }\arrowvert\rho (t)\arrowvert x_{1}\!-\!x_{1}^{%
\prime },x_{2}\!-\!x_{2}^{\prime }\rangle  \nonumber \\
&\times &\exp \left( -i2\vec{k}\cdot \vec{x}^{\prime }\right) .  \label{W}
\end{eqnarray}%
We note that Wigner function, (\ref{W}), is defined over the enlarged set $(%
\mathcal{Z}\oplus \mathcal{Z}_{2})$ with the natural prescription 
\[
\langle \vec{x}\arrowvert\rho (t)\arrowvert\vec{x}^{\prime }\rangle =0, 
\]%
which is true if some index $x_{j}\in \mathcal{Z}_{2}$ because the Wannier
basis is on the field of $\mathcal{Z}$.

Our definition of Wigner function satisfies the fundamental marginal
conditions \cite{Wigner}:%
\begin{eqnarray*}
\int\int\limits_{-\pi }^{\pi }d\vec{k}\ W(\vec{k},\vec{x},t) &=&\langle \vec{%
x}\arrowvert\rho (t)\arrowvert\vec{x}\rangle \geq 0 \\
\sum_{\vec{x}\in (\mathcal{Z}\oplus \mathcal{Z}_{2})}W(\vec{k},\vec{x},t)
&=&\langle \vec{k}\arrowvert\rho (t)\arrowvert\vec{k}\rangle \geq 0 \\
\sum_{\vec{x}\in (\mathcal{Z}\oplus \mathcal{Z}_{2})}\int\int\limits_{-\pi
}^{\pi }d\vec{k}\ W(\vec{k},\vec{x},t) &=&1.
\end{eqnarray*}

In addition, from the discrete Fourier transform we can obtain the \textit{%
inverse} relation on the field of integers ($\mathcal{Z}$):%
\begin{eqnarray*}
\langle s_{1},s_{2}\arrowvert\rho (t)\arrowvert s_{1}^{\prime
},s_{2}^{\prime }\rangle \! &=& \!\int\int\limits_{-\pi }^{\pi }d\vec{k}\
W(k_{1},k_{2},\frac{s_{1}+s_{1}^{\prime }}{2},\frac{s_{2}+s_{2}^{\prime }}{2}%
,t) \\
&\times &\exp [i\vec{k}\cdot (\vec{s}-\vec{s}^{\prime })].
\end{eqnarray*}%
Note that this inverse relation demands the necessity of a quasi-pdf on the
enlarged field of $(\mathcal{Z}\oplus \mathcal{Z}_{2})$. Carrying on the
calculation, from Eq.(\ref{W}) we arrive to 
\begin{eqnarray}
W(\vec{k},\vec{x},t)\! &=& \!\frac{e^{-2t_{D}}}{4\pi ^{2}}%
\!\!\!\!\!\!\sum\limits_{\{\alpha ,\beta ,q,n_{2},n_{3},n_{5}\}\in \emph{Z}%
}\!\!\!\!\!\!\!\!(-1)^{2x_{1}+2x_{2}}  \nonumber \\
\!\!\!\!\!\!\!\! &\times &\!J_{2x_{1}+2\alpha -q}(\!-2t_{\Omega }\sin
k_{1}\!)J_{2x_{2}+2\beta +q}(\!-2t_{\Omega }\sin k_{2}\!)  \nonumber \\
\!\!\!\!\!\!\!\! &\times &\!
(-1)^{q}(-1)^{n_{2}+n_{3}}I_{n_{2}}(t_{D})I_{n_{3}}(t_{D})I_{n_{5}}(t_{D}) 
\nonumber \\
\!\!\!\!\!\!\!\! &\times & \!I_{n_{2}+n_{5}-\alpha
}(t_{D})I_{n_{3}+n_{5}+\beta }(t_{D})I_{n_{2}+n_{3}+n_{5}-q}(t_{D}) 
\nonumber \\
\!\!\!\!\!\!\!\! &\times & \!\exp [iq(k_{1}-k_{2})],  \label{Wg}
\end{eqnarray}%
here we have used the IC (\ref{IC1}). The present definition is equivalent
to the one proposed from \textit{phase-point operators} in the reference 
\cite{Hinarejos}. In the case $D=0$ we obtain the well known non-dissipative
description for a tight-binding%
\[
W(\vec{k},\vec{x},t)_{D=0}=\frac{1}{4\pi ^{2}}J_{2x_{1}}(2t_{\Omega }\sin
k_{1})\ J_{2x_{2}}(2t_{\Omega }\sin k_{2}), 
\]%
this solution represents two-independent particles evolving with a unitary
transformation \cite{mm}.

Equation (\ref{Wg}) can be used to detect whether a state in phase-space has
a pure quantum character. This goal can be achieved by looking if $W(\vec{k},%
\vec{x};t)$ is negative or not. Therefore, the total negative volume in
phase-space can be measured by the (positive) function:%
\begin{equation}
\mathcal{V}(t)\!=\!\!\!\!\sum_{\vec{x}\in (\mathcal{Z}\oplus \mathcal{Z}%
_{2})}\int\int\limits_{-\pi }^{\pi }d\vec{k}\ \left[ \left\vert W(\vec{k},%
\vec{x},t)\right\vert -W(\vec{k},\vec{x},t)\right]  \label{Volum}
\end{equation}%
We note that in a situation where the classical regime dominates this
function vanishes, on the contrary in a quantum regime this function is
larger than zero. Then, this measure could be used to show the quantum to
classical transition. We can compare this result with the characteristic
time-scale --for the maximum of coherence-- that we have found using
different measures in section IV. 
\begin{figure}[t]
\centering
\includegraphics[width=0.4 \columnwidth,clip]{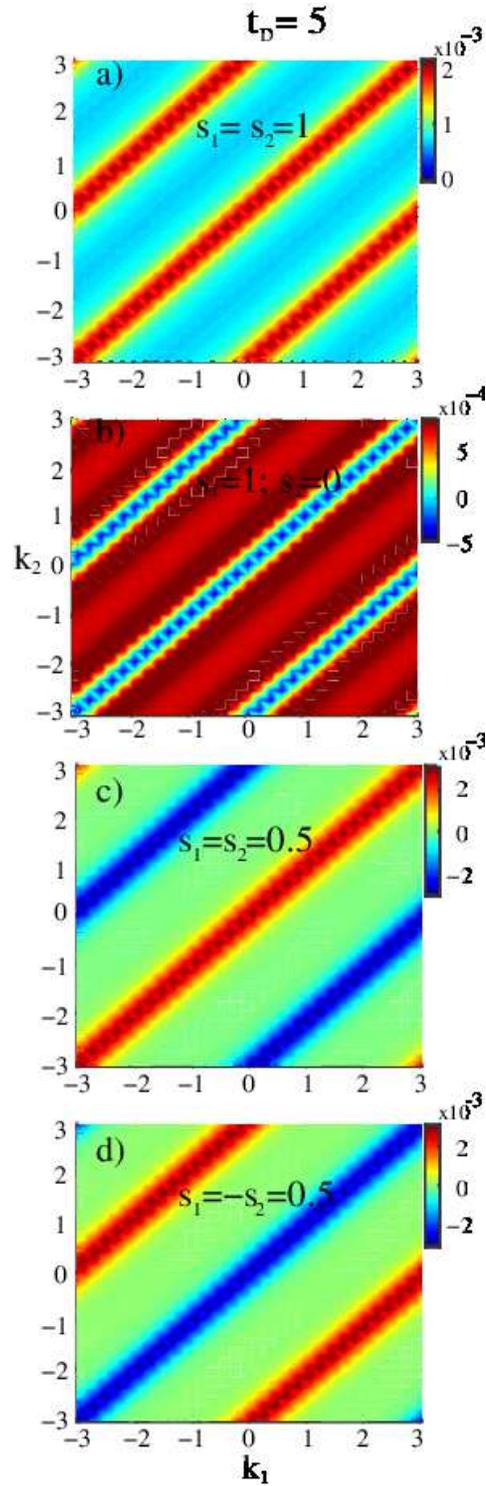}
\caption{(Color online) Wigner quasi pdf for two particles with IC as in (%
\protect\ref{IC1}) in the Fourier plane $\{k_{1},k_{2}\}$, for time $%
t_{D}=2Dt=5$. The negative domains are shown in dark blue. (a), (b) Wigner
function for the case: $\{s_{1}=-s_{2}=1$, $s_{1}=s_{2}=1\}$ (integers) and
(c), (d) $\{s_{1}=s_{2}=0.5$, $s_{1}=-s_{2}=0.5\}$ (semi-integers).}
\label{fig7}
\end{figure}

In figure \ref{fig7} we show several portraits of this pseudo pdf, in fact,
it is clear to identify the domain where the Wigner function is negative. In
particular, we plot $W(\vec{k},\vec{x},t)$ in the $k_{1},k_{2}$ plane for
the cases $s_{1}=s_{2}$ and $s_{1}=-s_{2}$. The plots show a strip structure
in the Brillouin zone and also the symmetry of the Wigner function (mirror
reflection on the plane $k_{1}=k_{2}$). The blue regions (color online)
correspond to negative values of the Wigner function. In general, we can
propose to use $W(\vec{k},\vec{x},t)$ to point out the quantum to classical
transition as a function of the dissipative parameter $r_{D}\equiv \frac{2D}{%
\Omega /\hbar }$ and the dimensionless time $t^{\prime }=t_{\Omega }$.

\begin{figure}[t]
\centering
\includegraphics[width=0.5 \columnwidth,clip]{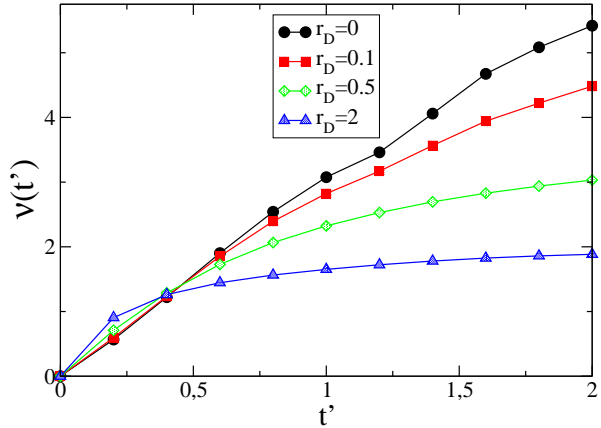}
\caption{(Color online) Absolute value of the negative volume of the Wigner
function for two particles as a function of $t^{\prime }=\frac{\Omega t}{%
\hbar }$. Here we have used the same parameters as in figure \protect\ref%
{fig5}, then the quantum character of the correlations is \textit{indicated}
by the non-zero value of this volume $\mathcal{V}(t)$.}
\label{fig8}
\end{figure}

In figure \ref{fig8} we show the absolute value of the negative volume $%
\mathcal{V}(t)$ as a function of $t^{\prime }=\frac{\Omega t}{\hbar }$, and
for different values of $r_{D}$. From this plot we reach to the conclusion
that for times $0<t^{\prime }<\tau _{c}^{\prime }$ the negative volume is
larger for the case $T\neq 0$ than with respect to the case $T=0$,
indirectly indicating the quantum character of the correlations. This
behavior is similar to the one that we got analyzing the QC (see Section
4.D). We want remark that $\tau _{c}^{\prime }$ is similar to the
scaling-time obtained from QC (see Fig. \ref{fig5}), and $\tau _{M}$ from
the GQD see Fig. \ref{fig8New}. 

From Eq.(\ref{Wg}) we can also see that if $\Omega =0$ the expression
simplify notably: 
\begin{eqnarray*}
W(\vec{k},\vec{x},t)_{\Omega =0} &=& \frac{e^{-2t_{D}}}{4\pi ^{2}}%
\sum\limits_{\{q_{1},q_{2}\}\in \emph{Z}%
}(-1)^{q_{1}+q_{2}}J_{q_{1}+x_{1}}(2Dt) \\
&\times &\!J_{q_{2}-x_{2}}(2Dt)I_{q_{2}+q_{1}}(2Dt\cos \left(
k_{2}-k_{1}\right) ) \\
&\times &\!I_{q_{1}-q_{2}}(2Dt\cos \left( k_{2}\!-\!k_{1}\right)
)I_{n_{2}+n_{3}+n_{5}-q}(t_{D}) \\
&\times &\!\exp [iq(k_{1}-k_{2})].
\end{eqnarray*}%
Then, it is possible to check that Wigner function has an interference
pattern even in the high dissipative regime $r_{D}\gg 1$, see figure \ref%
{fig7}. It is easy to see that in the case $\Omega =0$ it is also possible
to find phase-space domains where this quasi-pdf is negative.

\section{Quantum Correlations}

Different methods for characterizing the quantum and classical parts of
correlations is an active topics in quantum information theory \cite{nielsen}%
. An important part of the quantum information community consider the 
\textit{Quantum Discord} (QD)\ as a suitable measure of quantum correlations 
\cite{Discord}. Nevertheless there are some criticisms on this measure
because QD is not contractive under general local operations and therefore
should not be regarded as a strict measure for that purpose. Despite this
issue, and in order to compute an analytical expression for characterizing
correlations in our system, we have to argue in favor of QD \cite{Modi}.
Related with the QD is the \textit{geometric quantum discord} (GQD) \cite%
{GQD}. In particular, we use here, for such purpose, the GQD which is easier
to calculate than the QD (this measure involves an optimization procedure),
and it has been proved to be a necessary and sufficient condition for
non-zero QD \cite{GQD}.

\subsection{Geometric quantum discord of bipartite states.\textit{\ }}

The geometric measure of \textit{quantum discord} (GQD) has been defined as%
\begin{equation}
D_{G}\left( \rho \right) =\min_{\chi \in \Omega _{0}}\left\vert \left\vert
\rho -\chi \right\vert \right\vert ^{2},  \label{GQD}
\end{equation}%
where $\Omega _{0}$ denotes the set of zero-discord states and $\left\vert
\left\vert X-Y\right\vert \right\vert ^{2}=\mbox{Tr}\left( X-Y\right) ^{2}$
is the square norm in the Hilbert-Schmidt space. The lower bound of the GQD
can be calculated using that the density operator on a bipartite system
belonging to $H^{a}\otimes H^{b}$, with $\dim H^{a}=m$ and $\dim H^{b}=n$,
can be written in the form \cite{GQD,lowerGQD,3GQD,Rau}:%
\begin{eqnarray}
\rho &=&\frac{1}{mn}\left( \mathbf{I}_{m}\otimes \mathbf{I}_{n}+\sum_{i}x_{i}%
\tilde{\lambda}_{i}\otimes \mathbf{I}_{n}+\sum_{j}y_{j}\mathbf{I}_{m}\otimes 
\tilde{\lambda}_{j}\right.  \label{GQD1} \\
&&+\left. \sum_{j}t_{ij}\tilde{\lambda}_{i}\otimes \tilde{\lambda}%
_{j}\right) ,  \nonumber
\end{eqnarray}%
where $\tilde{\lambda}_{i},i=1,\cdots ,m^{2}-1$ and $\tilde{\lambda}%
_{j},j=1,\cdots ,n^{2}-1$ are the generators of $SU(m)$ and $SU(n)$
respectively, satisfying $\mbox{Tr}\left( \tilde{\lambda}_{i}\tilde{\lambda}%
_{j}\right) =2\delta _{ij}$. In this expression the vectors $\vec{x}\in 
\emph{R}^{m^{2}-1}$ and $\vec{y}\in \emph{R}^{n^{2}-1}$ of the subsystems $A$
and $B$ are given by: 
\begin{eqnarray*}
x_{i} &=&\frac{m}{2}\mbox{Tr}\left( \rho \tilde{\lambda}_{i}\otimes \mathbf{I%
}_{n}\right) =\frac{m}{2}\mbox{Tr}\left( \rho _{A}\tilde{\lambda}_{i}\right)
\\
y_{j} &=&\frac{m}{2}\mbox{Tr}\left( \rho \mathbf{I}_{m}\otimes \tilde{\lambda%
}_{j}\right) =\frac{n}{2}\mbox{Tr}\left( \rho _{B}\tilde{\lambda}_{j}\right)
,
\end{eqnarray*}%
and the correlation matrix $T\equiv \left[ t_{ij}\right] $ is given by 
\[
T\equiv \left[ t_{ij}\right] =\frac{mn}{4}\mbox{Tr}\left( \rho \tilde{\lambda%
}_{i}\otimes \tilde{\lambda}_{j}\right) . 
\]

The lower bound of the GQD can be written as: 
\begin{equation}
D_{G}\left( \rho \right) \geq \frac{2}{m^{2}n}\left( \left\vert \left\vert 
\vec{x}\right\vert \right\vert ^{2}+\frac{2}{n}\left\vert \left\vert
T\right\vert \right\vert ^{2}-\sum_{i=1}^{m-1}\eta _{i}\right) ,
\label{GQD2}
\end{equation}%
where $\eta _{i},i=1,2,\cdots ,m^{2}-1$ are eigenvectors of the matrix $%
\left( \vec{x}\vec{x}^{t}+\frac{2}{n}TT^{t}\right) $ arranged in
non-increasing order \cite{lowerGQD}.

For our present model ($2$ particles in an infinite lattice), we need to
introduce a bipartition on the lattice to study the GQD (a similar \textit{%
bipartition} was used for a $1$-particle Hamiltonian system in Ref. \cite%
{physA-2}), therefore introducing a bipartition we will end with a \textit{%
qutrit-qutrit system. }In our case $m=n=3$ and we can use the $SU(3)$
representation for calculating the GQD \cite{3GQD} in terms of the Gell-Mann
matrices $\tilde{\lambda}_{j}$.

\subsection{The mirror bipartition for a $2$ particles system}

In our case $\rho \left( t\right) $ has a diffusion-like time-behavior
limited by the quantum unitary evolution (see Eq. \ref{rhog}). Then, our
calculation would consist in assuming that for a fixed time
\textquotedblleft $t$\textquotedblright\ the DQW has evolved in a finite
domain supported by the basis of generators $SU(M_{t}),$ with $M_{t}\gg 1$
(see \ref{GQD1}). Our approximation consists in calculating the GQD under
the $SU(3)$ projection neglecting all no-mirror effects in the infinite
lattice using (\ref{GQD2}). A similar procedure was used for a spin system
under the $SU\left( 2\right) $ projection \cite{Zanardy}

\begin{figure}[t]
\centering
\includegraphics[height=1.8cm,width=8cm]{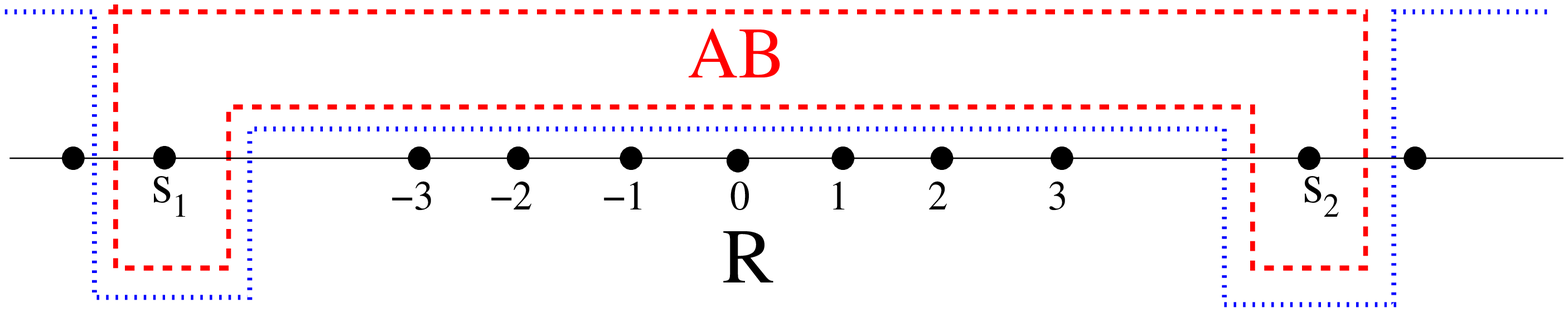}
\caption{Graphical representation of the bipartition on the lattice, the
sites $s_{1}=-s$ and $s_{2}=s$ are part of the subset $AB$, and the
remaining sites in the lattice is the complement subset $R$. The set $AB$
represents a qutrit-qutrit system. Then particles $1$ and $2$ can be either
in the set $AB$ or in $R$.}
\label{fig7New}
\end{figure}

In figure \ref{fig7New} we show the mirror bipartition that we are going to
use for the present $2$-body system, i.e., we trace out sites on the rest of
the lattice, keeping only two sites $\pm s$ in order to define a three-level
system; this bipartition allows us to calculate the GQD $D_{G}^{\left(
s\right) }\left( \rho _{AB}\right) .$ In other words, the density matrix $%
\rho _{AB}$ corresponds to the subset $AB$, that is: particles $1$ or $2$
can be at sites $\pm s$ or in the complement subset $R$ (the rest of the
lattice). Then it is clear that using this \textit{bipartition} we have
built up a couple qutrit-qutrit using the sites $\pm s$ on the lattice.

In order to trace out sites $s^{\prime }\neq \pm s$ note that for \textit{one%
} particle we can associate the kets 
\begin{eqnarray*}
\arrowvert A\rangle &\leftrightarrow &\arrowvert s\rangle \\
\arrowvert B\rangle &\leftrightarrow &\arrowvert-s\rangle \\
\arrowvert\phi \rangle &\leftrightarrow &\arrowvert s^{\prime }\rangle ,\
s^{\prime }\neq \pm s.
\end{eqnarray*}%
For $2$-particles the ket $\arrowvert s_{1},s_{2}\rangle $ can be written in
the form 
\begin{equation}
\arrowvert s_{1},s_{2}\rangle =\arrowvert\alpha \beta \rangle \otimes %
\arrowvert R\rangle ,  \label{GQD22}
\end{equation}%
where $\left\{ \alpha ,\beta \right\} \in \left\{ A,B,\phi \right\} $, and $%
R\ $is the complement, i.e., the set of sites, which are different from $\pm
s$.

Using (\ref{GQD22}) in our expression for the $2$-body density matrix (\ref%
{rhog}) we can calculate analytically the density matrix $\rho _{AB}$ for
the bipartition of figure\ \ref{fig7New}. This matrix turns to be a reduced $%
\left( 9\times 9\right) $ density matrix, i.e., it can be written in the
ordered basis:%
\[
\{\arrowvert AA\rangle ,\arrowvert AB\rangle ,\arrowvert A\phi \rangle ,%
\arrowvert BA\rangle ,\arrowvert BB\rangle ,\arrowvert B\phi \rangle ,%
\arrowvert\phi A\rangle ,\arrowvert\phi B\rangle ,\arrowvert\phi \phi
\rangle \} 
\]

In order to calculate the lower bound of the GQD we use the total mirror
contribution for the GQD, which is defined as 
\begin{equation}
D_{G}^{T}\left( \rho _{AB}\right) =\frac{d}{d-1}\sum_{s=1}^{\infty
}D_{G}^{\left( s\right) }\left( \rho _{AB}\right) ,  \label{GQD3}
\end{equation}%
where $D_{G}^{\left( s\right) }\left( \rho _{AB}\right) $ corresponds to (%
\ref{GQD2}) for a fixed value of $s$, here $d/(d-1)$ (which in our case
turns to be $3/2$) is a normalization factor. In fact the GQD can also be
defined through alternatives norms \cite{Paula}, here we used the norm$-2$
from references \cite{normGQD}, which by the way is equivalent to the
expression in \cite{Rau}.

\begin{figure}[t]
\centering
\includegraphics[width=0.5 \columnwidth,clip]{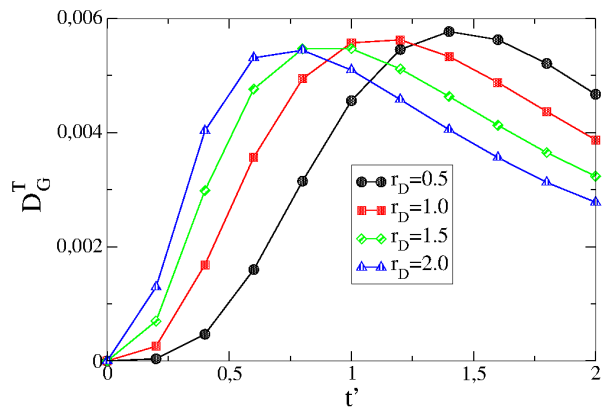}
\caption{Geometrical quantum discord calculated from the bipartition of
figure \protect\ref{fig7New}. The function $D_{G}^{T}\left( \protect\rho %
_{AB}\right) $ takes into account all the mirror contributions as defined in
(\protect\ref{GQD3}). This function is associated to a \textit{qutrit-qutrit}
system and represents the quantum correlations (bath-generated) between the
two particles. As noted there is a characteristic time $\protect\tau _{M}$
when these correlations are maxima.}
\label{fig8New}
\end{figure}

Then the value $D_{G}^{\left( s\right) }\left( \rho _{AB}\right) $ measures
the quantum correlation between particles $1$ and $2$ to be confined at
sites $\pm s.$ Using the dimensionless parameter $r_{D}=2D/(\Omega /\hbar )$
and the time $t^{\prime }=(\Omega /\hbar )t$ we have plotted $%
D_{G}^{T}\left( \rho _{AB}\right) $ as a function of time, for different
values of the temperature of the bath ($D\equiv \Gamma ^{2}k_{B}T/\hbar $).
In figure \ref{fig8New} we show the GQD (lower bound given by \ref{GQD2})
for three values of $r_{D}$, from this plot it is possible to see that the
correlations induced between the particles are in fact of quantum nature
because $D_{G}^{T}\left( \rho _{AB}\right) >0$ for almost all $t>0$ (for the
case $r_{D}=0.5$ and at short times --apart form numerical errors-- it seems
to be negative). Then, several important conclusions can be drawn. 

\textit{First}: The bath-induced quantum correlation is time dependent
showing a non-monotonic behavior, with a maximum at a characteristic
time-scale $\tau _{M}$ that depends on the dissipative parameter $r_{D}$.
After this time quantum correlations are wiped\ out by the dissipation.

\textit{Second}: If the temperature of the bath decreases the characteristic
time $\tau _{M}$ is delayed. If $r_{D}=0$ the GQD vanishes at all times.

\textit{Third}: The bath-induced quantum correlation would vanish
monotonically at long-time as a result of the dissipation (we could not run
the long-time behavior because we have numerical errors). The behavior of $%
D_{G}^{T}\left( \rho _{AB}\right) $ as a function of time $t^{\prime }$ has
a response (with a maximum) with a similar time-scale as the one from the QC
and the negative volume of the Wigner function.

\section{Conclusions.}

We have analyzed two free spinless particles --in a 1D regular lattice--
interacting with a common thermal phonon bath $\mathcal{B}$. We have solved
analytically the associated QME for the two (initially uncorrelated)
distinguishable particles to characterize the quantumness of the
correlations induced by $\mathcal{B}$.

Some measures like spatiotemporal correlation function $\mathcal{C}_{(1,2)}$%
, quantum coherence $\mathcal{G}$, purity, entropy, quantum mutual
information have been analyzed showing a high degree of coherence between
the two particles. Then, pointing out that this result is in fact induced by
the common bath, despite the existing dissipation for temperatures different
from zero ($D\equiv \Gamma ^{2}k_{B}T/\hbar $ where $T$ is the temperature
of $\mathcal{B}$). The purity, quantum mutual information, and the quantum
coherence $\mathcal{G}$ has been used to find a characteristic time-scale to
show when the coherence has a maximum value. We have find a similar
time-scale value for all these measures.

All these results have also been supported by the analysis of a Wigner-like
function which shows domains (in a lattice phase-space) being negative. We
have calculated the (absolute value) of the negative volume $\mathcal{V}(t)$
of the Wigner pdf as a function of time $t^{\prime }=\frac{\Omega t}{\hbar }$
and for several values of dissipative parameter $r_{D}$. In fact the
negative volume also shows a characteristic time-scale in its behavior which
is in agreement with our calculations made with other quantum measures of
coherence as the purity, off-diagonal quantum coherence, and quantum mutual
information. Thus the negative volume \textit{may be} considered as an
indirect indicator of quantum properties.

As a criterion to check the quantum nature of the process we have also
calculate the GQD. To do this we introduce a bipartition in the lattice in
order to associate a qutrit-qutrit set from our $2$-particle system, thus we
showed that $D_{G}^{T}\left( \rho _{AB}\right) >0$ for some values of $r_{D}$
and $t^{\prime }>0$, which would indicate the quantum nature of the induced
correlations between the particles. In fact the time behavior of the $%
D_{G}^{T}\left( \rho _{AB}\right) $ is an increasing function of time
showing a maximum at a characteristic time $\tau _{M}$, which is of the
order of the time-scale that we have found analyzing the QC and the Wigner
function. After this time $\tau _{M}$ the behavior of the GQD is decreasing
approaching zero.

The profile of probability for two particles is ballistic for $D=0$ (the
closed system corresponds to the tight-binding model) and it starts to be
modified by the presence of temperature ($D>0$), showing a notable X-form
pattern in the presence of dissipation. In the case of larger values of the
dissipation parameter $r_{D}\equiv \frac{2D}{\Omega /\hbar }\gg 1$ this
structure is accentuated and additional interference patterns are also
observed as a result of the interaction with the thermal bath $\mathcal{B}$.

Our approach opens the possibility of performing analytical analysis on
important quantities related to quantum correlation measures in dissipative
bipartite systems \cite{physA-1,physA-2}, for example the quantum discord
for a qutrit-qutrit set as we have shown in this work. In addition from the
present approach it is possible to write down an equivalent QME for more
than two (free) bosonic or fermionic particles interacting with a common
bath, these results are of great interest in nanoscience and quantum
information theory, and will be presented in future contributions. Then, our
analysis could be of interest in experiments using --for example-- photonic
devices similar to the ones of Refs. \cite{peruzzo,perets}.

\textit{Acknowledgments.} M.O.C gratefully acknowledges support received
from CONICET, grant PIP 112-201501-00216 CO, Argentina.

\section*{Appendix A: Fock representation for Bosonic particles}

In this appendix, we prove the equivalence between the Fock representation
and the Wannier basis. In particular we will work out bosonic particles in
Fock's space and symmetric Wannier's states. A similar equivalent
demonstration can be done for fermionic particles. Let us define the
creation operators $C_{s_{1}}^{\dagger },C_{s_{2}}^{\dagger }$ acting on the
vacuum state $\arrowvert\phi \rangle $. Fock's basis can be denoted as%
\begin{equation}
C_{s_{1}}^{\dagger }C_{s_{2}}^{\dagger }\arrowvert\phi \rangle \!=\!%
\arrowvert\cdots ,0,\cdots ,1,0,\cdots ,1,0,\cdots ,\cdots ,0,\cdots \rangle
,  \label{Apen1}
\end{equation}%
i.e., it has been created two indistinguishable particles at sites $s_{1}$
and $s_{2}$. If these particles are bosonic the operators $%
C_{s_{1}}^{\dagger },C_{s_{2}}^{\dagger }$ and $C_{s_{1}},C_{s_{2}}$ satisfy
the relations%
\begin{equation}
\left[ C_{s_{1}},C_{s_{2}}^{\dagger }\right] =\delta _{s_{1},s_{2}};\ \left[
C_{s_{1}},C_{s_{2}}\right] =\left[ C_{s_{1}}^{\dagger },C_{s_{2}}^{\dagger }%
\right] =0.  \label{Apen2}
\end{equation}

Let the translation operator be defined as%
\begin{equation}
R=\sum_{s=-\infty }^{\infty }C_{s-1}^{\dagger }C_{s}  \label{Apen3}
\end{equation}%
Therefore, using (\ref{Apen2}) we can prove that $R$ translates
(individually) two indistinguishable particles 
\begin{eqnarray}
RC_{s_{1}}^{\dagger }C_{s_{2}}^{\dagger }\arrowvert\phi \rangle
&=&\sum_{s=-\infty }^{\infty }C_{s-1}^{\dagger }C_{s}C_{s_{1}}^{\dagger
}C_{s_{2}}^{\dagger }\arrowvert\phi \rangle  \nonumber \\
&=&\sum_{s=-\infty }^{\infty }C_{s-1}^{\dagger }\left( \delta
_{s,s_{1}}+C_{s_{1}}^{\dagger }C_{s}\right) C_{s_{2}}^{\dagger }\arrowvert%
\phi \rangle  \nonumber \\
&=&C_{s_{1}-1}^{\dagger }C_{s_{2}}^{\dagger }\arrowvert\phi \rangle 
\nonumber \\
&+&\sum_{s=-\infty }^{\infty }C_{s-1}^{\dagger }C_{s_{1}}^{\dagger }\left(
\delta _{s,s_{2}}+C_{s_{2}}^{\dagger }C_{s}\right) \arrowvert\phi \rangle 
\nonumber \\
&=&C_{s_{1}-1}^{\dagger }C_{s_{2}}^{\dagger }\arrowvert\phi \rangle
+C_{s_{2}-1}^{\dagger }C_{s_{1}}^{\dagger }\arrowvert\phi \rangle ,
\label{Apen4}
\end{eqnarray}%
here we have used that $C_{s}\arrowvert\phi \rangle =0.$ In a similar way,
noting that 
\begin{equation}
R^{\dagger }\!\!=\!\!\sum_{s=-\infty }^{\infty }\left( C_{s-1}^{\dagger
}C_{s}\right) ^{\dagger }\!\!=\!\!\!\!\sum_{s=-\infty }^{\infty
}C_{s}^{\dagger }C_{s-1}=\sum_{s^{\prime }=-\infty }^{\infty }C_{s^{\prime
}+1}^{\dagger }C_{s^{\prime }},  \label{Apen5}
\end{equation}%
it is simple to prove that%
\begin{equation}
R^{\dagger }C_{s_{1}}^{\dagger }C_{s_{2}}^{\dagger }\arrowvert\phi \rangle
=C_{s_{1}+1}^{\dagger }C_{s_{2}}^{\dagger }\arrowvert\phi \rangle
+C_{s_{2}+1}^{\dagger }C_{s_{2}}^{\dagger }\arrowvert\phi \rangle .
\label{Apen6}
\end{equation}

For the calculation of the infinitesimal Kossakoswki-Lindbland generator it
is important to know the action of the operator $R^{\dagger }R$. To
calculate this operator, we use the definition of $R^{\dagger },R$ and (\ref%
{Apen2}), then we get%
\begin{eqnarray}
RR^{\dagger } &=&\sum_{s=-\infty }^{\infty }C_{s-1}^{\dagger
}C_{s}\sum_{s^{\prime }=-\infty }^{\infty }C_{s^{\prime }+1}^{\dagger
}C_{s^{\prime }}  \nonumber  \label{Apen7} \\
&=&\sum_{s^{\prime }=-\infty }^{\infty }C_{s^{\prime }}^{\dagger
}C_{s^{\prime }}+\sum_{s=-\infty }^{\infty }\sum_{s^{\prime }=-\infty
}^{\infty }C_{s-1}^{\dagger }C_{s^{\prime }+1}^{\dagger }C_{s}C_{s^{\prime
}}.  \nonumber
\end{eqnarray}%
Therefore 
\begin{eqnarray}
RR^{\dagger }C_{s_{1}}^{\dagger }C_{s_{2}}^{\dagger }\arrowvert\phi \rangle
&=&2C_{s_{1}}^{\dagger }C_{s_{2}}^{\dagger }\arrowvert\phi \rangle
+C_{s_{1}-1}^{\dagger }C_{s_{2}+1}^{\dagger }\arrowvert\phi \rangle 
\nonumber \\
&+&C_{s_{1}+1}^{\dagger }C_{s_{2}-1}^{\dagger }\arrowvert\phi \rangle .
\label{Apen77}
\end{eqnarray}

In a similar way we can also prove that\ $RR^{\dagger }=R^{\dagger }R$,
therefore, for boson particles $R$ and $R^{\dagger }$\ commute, i.e., $\left[
R,R^{\dagger }\right] =0.$

Now we are going to show that the same algebra can be obtained using a
symmetrized Wannier's basis. This is an important result for calculating the
infinitesimal generator. Let the symmetric Wannier vector state (for two
particles) be written in the form%
\begin{equation}
\arrowvert s_{1},s_{2}\rangle _{S}=\frac{1}{\sqrt{2}}\left[ \arrowvert %
s_{1},s_{2}\rangle +\arrowvert s_{2},s_{1}\rangle \right] .  \nonumber
\end{equation}%
In order to compare both algebras --in Fock and Wannier spaces-- we need to
know the action of translation operators in the symmetric Wannier
representation. Then, from (\ref{Apen6}) we can define the action of the
translation operator $T_{12}$ for indistinguishable particles as%
\begin{equation}
T_{12}\arrowvert s_{1},s_{2}\rangle _{S}=\arrowvert s_{1}-1,s_{2}\rangle
_{S}+\arrowvert s_{1},s_{2}-1\rangle _{S}  \label{T12}
\end{equation}%
\begin{equation}
T_{12}^{\dagger }\arrowvert s_{1},s_{2}\rangle _{S}=\arrowvert %
s_{1}+1,s_{2}\rangle _{S}+\arrowvert s_{1},s_{2}+1\rangle _{S}.
\label{T12mas}
\end{equation}%
These operators produce the same state as when $R$ and $R^{\dagger }$ are
applied to Fock's vector state. To end this appendix we can calculate here
the action of $T_{12}^{\dagger }T_{12}$ on a symmetric Wannier's state, from
(\ref{T12}) and (\ref{T12mas}) we get%
\[
T_{12}^{\dagger }T_{12}\arrowvert s_{1},\!s_{2}\rangle _{S}\!=\!2\arrowvert %
s_{1},\!s_{2}\rangle _{S}+\arrowvert s_{1}-1,\!s_{2}+1\rangle _{S}+%
\arrowvert s_{1}+1,\!s_{2}-1\rangle _{S} 
\]%
with $\left[ T_{12}^{\dagger },T_{12}\right] =0$. Therefore, the action of $%
T_{12}^{\dagger }T_{12}$ on a symmetric Wannier state is equivalent to the
action of $R^{\dagger }R$ on a Bosonic Fock state (\ref{Apen77}). Similar
calculation can be done for fermionic particles.

\section*{Appendix B: The two-body reduced density matrix from a \textit{pure%
} interacting infinitesimal generator}

Consider the infinitesimal generator for two distinguishable DQW in the
Fourier representation (\ref{F22}). If we want to solve a pseudo-density
matrix $\Pi (t)$ taking into account \textit{only} the bath-mediated
interacting term; then, the evolution equation will be%
\begin{equation}
\frac{d}{dt}\!\left\langle k_{1},\!k_{2}\left \vert \Pi (t)\right\vert
k_{1}^{\prime },\!k_{2}^{\prime }\right\rangle \!\!=\!\!\mathcal{I}\!\left(
k_{1},k_{2},\!k_{1}^{\prime },\!k_{2}^{\prime }\right)\! \left\langle
k_{1},k_{2}\left\vert \Pi (t)\right\vert k_{1}^{\prime },\!k_{2}^{\prime
}\right\rangle \!,  \label{B1}
\end{equation}%
with%
\begin{eqnarray}
\mathcal{I}\left( k_{1},k_{2},k_{1}^{\prime },k_{2}^{\prime }\right)
\!\!&=&\!\!2D[\mathbf{C}\left( k_{1}\!,k_{2}^{\prime }\right) \!+\!\mathbf{C}%
\left( k_{2}\!,k_{1}^{\prime }\right) \!-\!\mathbf{C}\left(
k_{1}\!,k_{2}\right)  \nonumber \\
&-&\!\!\mathbf{C}\left( k_{1}^{\prime }\!,k_{2}^{\prime }\right) ].
\label{B11}
\end{eqnarray}

If the IC is $\Pi (0)=\arrowvert s_{1}^{0},s_{2}^{0}\rangle \langle
s_{1}^{0},s_{2}^{0}\arrowvert$, the solution will be given by the Fourier
antitransform%
\begin{eqnarray*}
\left\langle s_{1},s_{2}\left\vert \Pi (t)\right\vert s_{1}^{\prime
},s_{2}^{\prime }\right\rangle \! &=&\!\left( 2\pi \right) ^{-4}\int \int
\int \int dk_{1}\!\!\ dk_{1}^{\prime }\!\!\ dk_{2}\!\!\ dk_{2}^{\prime } \\
&\times &\!e^{ik_{1}(s_{1}-s_{1}^{0})}e^{ik_{1}^{\prime }(-s_{1}^{\prime
}+s_{1}^{0})} \\
&\times &\!e^{ik_{2}(s_{2}-s_{2}^{0})}e^{ik_{2}^{\prime }(-s_{2}^{\prime
}+s_{2}^{0})}e^{\mathcal{I}\left( k_{1},k_{2},k_{1}^{\prime }k_{2}^{\prime
}\right) t}\!,
\end{eqnarray*}%
using $e^{x\cos \theta }=\sum_{n=-\infty }^{\infty }I_{n}(x)e^{in\theta }$,
and after some algebra (noting that $I_{n}(-x)=(-1)^{n}I_{n}(x)$ and using (%
\ref{bessel1})-(\ref{bessel2})) we get%
\begin{eqnarray}
\left\langle s_{1},s_{2}\left\vert \Pi (t)\right\vert s_{1}^{\prime
},s_{2}^{\prime }\right\rangle \! &=& \!\delta _{s_{1},s_{2},s_{1}^{\prime
},s_{2}^{\prime }}\sum_{n_{4}\in \mathcal{Z}}I_{-s_{2}^{\prime
}+s_{2}^{0}-n_{4}}(t_{D})  \nonumber \\
&\times & \!I_{-s_{1}^{\prime
}+s_{1}^{0}-n_{4}}(t_{D})I_{s_{2}-s_{2}^{0}-s_{1}^{\prime
}+s_{1}^{0}+n_{4}}(t_{D})  \nonumber \\
&\times & \!I_{n_{4}}(t_{D}),  \label{sol1}
\end{eqnarray}%
from this result it is simple to see that this solution is normalized $%
\mbox{Tr}\left[ \Pi (t)\right] =1.$ Two interesting conclusions can also be
drawn from (\ref{sol1}): the first is concerning the Purity of the two-body
system, and the second is about the \textit{one-body }reduced density matrix.

1) From the solution (\ref{sol1}), and after some algebra, we can calculate
the Purity associated to the pseudo density matrix

\begin{eqnarray*}
\mathcal{P}_{Q}(t) &=&\mbox{Tr}\lbrack \Pi (t)^{2}] \\
&=&\sum_{n=-\infty }^{\infty }\left( -1\right) ^{n+s_{1}^{0}-s_{2}^{0}}\
I_{n+s_{1}^{0}-s_{2}^{0}}^{4}(2t_{D}) \\
&=&\sum_{m=-\infty }^{\infty }\left[ \left( -1\right) ^{m}\ I_{m}^{2}(2t_{D})%
\right] ^{2}\equiv \sum_{m=-\infty }^{\infty }\beta _{m}^{2}.
\end{eqnarray*}%
Therefore we conclude that \textquotedblleft the
eigenvalues\textquotedblright\ (for fixed time $t$) of this two-body pseudo
density matrix are not necessarily positive $\beta _{m}=\left( -1\right)
^{m}\ I_{m}^{2}(2t_{D})$. \ This result comes from the fact that $\mathcal{I}%
\left( k_{1},k_{2},k_{1}^{\prime }k_{2}^{\prime }\right) $ is not a complete
infinitesimal generator. The present analysis only helps to give us light
into the meaning of the bath-mediated interaction term.

2) Tracing over one particle, say label $``2``$, we get%
\[
\langle s_{1}|\Pi ^{(1)}(t)|s_{1}^{\prime }\rangle )\!=\!\langle s_{1}%
\arrowvert\mbox{Tr}_{2}[\Pi (t)]\arrowvert s_{1}^{\prime }\rangle
\!=\!(\!-1\!)^{s_{1}+s_{1}^{0}}\delta _{s_{1},s_{1}^{0}}\delta
_{s_{1},s_{1}^{\prime }}, 
\]%
then, this pseudo one-body solution does not evolve in time. This conclusion
is consistent with the interpretation of the bath-mediated interaction (\ref%
{B11}) that we have used to calculate the time evolution of the pseudo
density matrix $\Pi (t)$ in (\ref{B1}).

\end{document}